\documentclass[twocolumn,pre,showpacs]{revtex4}
\usepackage{amssymb}
\usepackage{amsfonts}
\usepackage{amsmath}
\usepackage{graphicx}

\begin{document}
\title{A generalized fluctuation relation for power-law distributions}
\author{Adri\'{a}n A. Budini}
\affiliation{Consejo Nacional de Investigaciones Cient\'{\i}ficas y T\'{e}cnicas
(CONICET), Centro At\'{o}mico Bariloche, Avenida E. Bustillo Km 9.5, (8400)
Bariloche, Argentina}
\date{\today}
\begin{abstract}
Strong violations of existing fluctuation theorems may arise in
nonequilibrium steady states characterized by distributions with power-law
tails. The ratio of the probabilities of positive and negative fluctuations
of equal magnitude behaves in an anomalous nonmonotonic way [H. Touchette
and E.G.D. Cohen, Phys. Rev. E \textbf{76}, 020101(R) (2007)]. Here, we
propose an alternative definition of fluctuation relation (FR) symmetry
that, in the power-law regime, is characterized by a monotonic linear
behavior. The proposal is consistent with a large deviation-like principle.
As example, it is studied the fluctuations of the work done on a dragged
particle immersed in a complex environment able to induce power-law tails.
When the environment is characterized by spatiotemporal temperature
fluctuations, distributions arising in nonextensive statistical mechanics
define the work statistics. In that situation, we find that the FR symmetry
is solely defined by the average bath temperature. The case of a dragged
particle subjected to a L\'{e}vy noise is also analyzed in detail.
\end{abstract}
\maketitle

\section{Introduction}

% 05.70.Ln, NonEquilibrium and irreversible thermodynamics
% 05.40.-a, Fluctuation phenomena, random processes, noise, and Brownian motion
% 02.50.-r Probability theory, stochastic processes, and statistics
% 05.70.-a, Thermodynamics

Fluctuation theorems have become a standard tool to characterize
nonequilibrium states \cite%
{gallavoti,Jarzynski,kurchan,spohn,crocks,maes,seifert}. Independently of
their specific formulation \cite{searles,vulpiani,mukamel,JarzynskiReport},
a common underlying ingredient is an assertion about the symmetries of the
fluctuations measures. For example, one of the most common formulation
establishes that the probability of positive and negative fluctuations of a
given variable differ by an exponential weight proportional to the
fluctuations magnitude. This FR symmetry has been confirmed in a wide class
of experimental setups \cite{PRLexper}. While most of the analysis focus on
variables such as entropy production and work (performed on a system), the
FR symmetries may also be valid for nonthermodynamic variables \cite%
{soodPituto}.

A Brownian particle dragged by a spring through a thermal environment is one
of the simpler arrangement where a FR symmetry can be theoretically
predicted \cite{mazonka,zon} and measured \cite{wangBrownianExp}. While the
work performed on the particle satisfy a standard or conventional FR, the
heat fluctuations follow and extended FR \cite{FTExtended}. The exponential
weight that relates the probability of positive and negative fluctuations
does not scale linearly with the heat variable. Similar deviations with
respect to a linear dependence have been found in the injected power to
systems driven by an external stochastic force \cite{cohenPoisson,
FTNonLinear}.

Conventional or extended FRs \cite{FTExtended,cohenPoisson, FTNonLinear} are
essentially valid when the probability distributions obey a large deviation
principle (LDP) \cite{touchette,sornette}. This formalism also provides a
solid basis for characterizing nonequilibrium states \cite{garrahan}.
Therefore, stronger violations of conventional FRs should arise when
power-law distributions \cite{sornette} determine the fluctuations
statistics. In fact, self-similar structures can not be studied in the
context of a standard large deviation theory \cite{sornette,touchette}.

One of the first analysis about the incompatibility of FRs with anomalous
(power-law distributed) fluctuations was done by Touchette and Cohen in Ref. 
\cite{cohenRapid}. In contrast with previous Langevin models \cite%
{mazonka,zon}, where the environment influence is taken into account through
a Gaussian white noise, they considered a stable L\'{e}vy white noise. It
was found that the ratio of the probabilities of positive and negative work
fluctuations of equal magnitude behaves in an anomalous nonlinear way,
developing a convergence to one for large fluctuations. Hence, negative
fluctuations of the work performed on the particle are just as likely to
happen as large positive work fluctuations of equal magnitude. This
non-usual property strongly departs from the Gaussian case, where the
validity of a standard FR implies that positive fluctuations are
exponentially more probable than negative ones.

A similar analysis on anomalous fluctuation properties was performed by
Chechkin and Klages \cite{klages} for the same kind of Langevin models. In
the (power-law) L\'{e}vy case the same conclusions were obtained. On the
other hand, it was shown that standard FRs remain valid in presence of
normal long time correlated fluctuations. The long-range correlations only
alter the time-speed of the large deviation functions (LDFs) \cite{touchette}%
. An analogous conclusion was obtained by Harris and Touchette \cite{harris}.

In Ref. \cite{BeckCohenFT} Beck and Cohen introduced an alternative FR that
arises by considering a superstatistical model \cite{beckQ,superstatistics},
where the particle environment develops (spaciotemporal) temperature
fluctuations. As is well known \cite{beckQ}, this dynamics leads to
power-law distributions arising in Tsallis nonextensive statistical
mechanics \cite%
{TsallisBook,Qlectures,qConstraints,stepestQ,QCentralLimit,LevyAsQ,QFokker}.
The alternative FR was derived by averaging temperature in a standard FR.
When the \textquotedblleft measured\textquotedblright\ work distribution
develops power-law tails \cite{BeckCohenFT}, a very complex expression that
does not has a clear physical meaning is obtained.

From different points of view, all quoted analysis \cite{klages,BeckCohenFT}
sustain the main conclusion of Ref. \cite{cohenRapid}, that is, FRs in
presence of power-law distributions acquire an anomalous (complicated)
structure whose origin can be linked to the incompatibility of self-similar
structures with a (standard) LDP. The main goal of this paper is to
introduce a generalized and alternative FR such that in presence of
power-law tails the symmetry between positive and negative fluctuations is
expressed through a linear dependence. Hence, a simple scheme for
understanding previous results \cite{cohenRapid,klages,harris,BeckCohenFT}
is obtained. Furthermore, we associate the proposed symmetry with a large
deviation-like principle. In a long time regime, it allows expressing the FR
through the symmetries of a set of LDFs associated to the probability
distribution and its characteristic function. As in the standard case \cite%
{spohn}, both functions are related by a Legendre-Fenchel transformation 
\cite{touchette}.

Similarly to the case of standard FR, the proposed probability symmetry does
not assume an underlying thermodynamic equilibrium, either extensive or
nonextensive. Nevertheless, the alternative FR symmetry adopts a very simple
structure when written in terms of a set of functions introduced in
nonextensive entropy formalism \cite{TsallisBook,Qlectures}. The relevance
of this stretched relation becomes evident when considering previous related
analysis \cite{BeckCohenFT}. Furthermore, it is expected that a FR
associated to a nonextensive (stochastic) thermodynamics may have a central
role for understanding intrinsic fluctuations in nanoscale systems \cite%
{NanoSuperFT}.

The paper is outlined as follows. In Sec. II we introduce the generalized FR
symmetry. In Sec. III we apply the alternative definition to a specific
physical system characterized by power-law tails. It is analyzed the
fluctuations of the work done on a dragged particle immersed in a complex
environment that develops (spatiotemporal) temperature fluctuations \cite%
{BeckCohenFT}. Furthermore, the L\'{e}vy model introduced in Ref. \cite%
{cohenRapid} is analyzed in detail. In Sec. IV we give the conclusions. In
Appendix A we develop a large deviation-like principle and characterize the
symmetries of the LDFs. These last results rely on a saddle-point
approximation performed in Appendix B.

\section{Generalized fluctuation relation}

For an arbitrary stochastic variable $x_{st},$ with probability distribution 
$p(x),$ a standard FR symmetry is defined by the relation $\ln
[p(x)/p(-x)]=\zeta x,$ or equivalently $p(\pm x)=p(\mp x)e^{\pm \zeta x},$
where $\zeta $ is a positive constant. Here, we propose the alternative FR
symmetry%
\begin{equation}
\frac{\lbrack p(x)]^{1-q}-[p(-x)]^{1-q}}{1-q}=\zeta x,  \label{FRLogIntro}
\end{equation}%
which in turn can be expressed as%
\begin{equation}
p(\pm x)=\{[p(\mp x)]^{1-q}\pm (1-q)\zeta x\}^{\frac{1}{1-q}}.
\label{FRProbaIntro}
\end{equation}%
The real parameter $q$ is related with the index of the power-law tails. Its
domain will be specified later on. The units of the constant $\zeta $ are $%
[1/x]^{2-q}.$ Nevertheless, in the standard case it is a conjugate variable
of $x,$ that is, its units are $[1/x].$ This discrepancy can be avoided by
writing%
\begin{equation}
\zeta =\frac{\zeta _{ph}}{\int_{-\infty }^{\infty }dx[p(x)]^{q}},
\label{zitaPhys}
\end{equation}%
where now $\zeta _{ph}$ play the role of a (physical) conjugate variable.
Notice that even with this parameter redefinition, the proposed symmetry can
be written solely in terms of $p(x).$ The generalized FR written in terms of 
$\zeta $ is called \textquotedblleft unnormalized scheme,\textquotedblright\
while in terms of $\zeta _{ph}$ \textquotedblleft normalized
scheme.\textquotedblright

The previous relations, Eqs. (\ref{FRLogIntro}) and (\ref{FRProbaIntro}),
strongly departs from the standard ones. Their structure is simplified by
writing them in terms of a set of functions introduced previously in the
context of Tsallis nonextensive statistical mechanics \cite{TsallisBook}. A
q-logarithm and q-exponential functions are defined respectively as%
\begin{equation}
\ln _{q}x\equiv \frac{x^{1-q}-1}{1-q},\ \ \ \ \ e_{q}^{x}\equiv \lbrack
1+(1-q)x]^{\frac{1}{1-q}},
\end{equation}%
jointly with the generalized q-product operation%
\begin{equation}
x\otimes _{q}y\equiv \lbrack x^{1-q}+y^{1-q}-1]^{\frac{1}{1-q}}.
\end{equation}%
The relevance of these definitions comes from the property that in the limit 
$q\rightarrow 1$ they recover the standard logarithm $[\ln _{1}x=\ln x],$
the exponential function $[\exp _{1}x=\exp x],$ and respectively the
standard product $[x\otimes _{1}y=xy].$ In terms of them, Eq. (\ref%
{FRLogIntro}) can be written as%
\begin{equation}
\ln _{q}[p(x)]-\ln _{q}[p(-x)]=\zeta x,  \label{FRLog}
\end{equation}%
while Eq. (\ref{FRProbaIntro}) becomes equivalent to 
\begin{equation}
p(-x)=p(x)\otimes _{q}e_{q}^{-\zeta x},\ \ \ \ \ \ p(x)=p(-x)\otimes
_{q}e_{q}^{\zeta x}.  \label{Prelation1}
\end{equation}%
Hence, the proposed FR can be read as a \textquotedblleft
deformation\textquotedblright\ of the standard ones. The parameter $q$
measures the degree of departure with respect to the standard case $%
(q\rightarrow 1).$ Notice that by using the properties $\ln _{q}(x\otimes
_{q}y)=\ln _{q}x+\ln _{q}y,$ and $e_{q}^{x+y}=e_{q}^{x}\otimes _{q}e_{q}^{y}$
\cite{TsallisBook}, the consistence between the previous two expressions
becomes evident.

The standard FR symmetry can also be written in terms of the characteristic
function $Z(\lambda )=\int_{-\infty }^{\infty }dxp(x)e^{-\lambda x},$ that
is, $Z(\lambda )=Z(-\lambda +\zeta ),$ which in turn implies $Z(\zeta )=1.$
Taking into account Eq. (\ref{FRProbaIntro}), it follows that here we must
consider the generalized expression%
\begin{equation}
Z_{q}(\lambda )\equiv \int_{-\infty }^{\infty }dx\{[p(x)]^{1-q}-(1-q)\lambda
x\}^{\frac{1}{1-q}},
\end{equation}%
which can be rewritten as 
\begin{equation}
Z_{q}(\lambda )=\int_{-\infty }^{\infty }dx[p(x)\otimes _{q}e_{q}^{-\lambda
x}].  \label{qGeneratingFunction}
\end{equation}%
By using the associative property $x\otimes _{q}(y\otimes _{q}z)=(x\otimes
_{q}y)\otimes _{q}z$ \cite{TsallisBook}, the generalized FR [Eq. (\ref{FRLog}%
) or (\ref{Prelation1})] implies the equivalent symmetry%
\begin{equation}
Z_{q}(\lambda )=Z_{q}(-\lambda +\zeta ),\ \ \ \ \ \ \ \ \ \ \ \ \ \
Z_{q}(\zeta )=1.  \label{FR_Generating}
\end{equation}%
Notice that this condition is exactly the same that defines the standard
case, $q\rightarrow 1.$

The expression (\ref{qGeneratingFunction}) naturally arises in nonextensive
statistical mechanics. Hence, it properties are well known \cite{TsallisBook}%
. $Z_{q}(\lambda )$ is the generating function of a kind of generalized
moments of $x_{st},$ which are calculated from powers of $p(x).$

In accordance with the normalized scheme [Eq. (\ref{zitaPhys})], we
introduce the definitions 
\begin{equation}
Z_{q}^{ph}(\lambda _{ph})\equiv Z_{q}(\lambda (\lambda _{ph})),\ \ \ \ \ \ \
\ \ \lambda =\frac{\lambda _{ph}}{\int_{-\infty }^{\infty }dx[p(x)]^{q}}.
\end{equation}%
Thus, Eqs. (\ref{zitaPhys}) and (\ref{FR_Generating}) allows us to write the
equivalent symmetry%
\begin{equation}
Z_{q}^{ph}(\lambda _{ph})=Z_{q}^{ph}(-\lambda _{ph}+\zeta _{ph}),\ \ \ \ \ \
\ \ \ \ \ \ \ \ Z_{q}^{ph}(\zeta _{ph})=1.
\end{equation}

\subsection{q-Gaussian distributions}

The relations (\ref{FRLog}) and (\ref{Prelation1}) define the generalized
FR. Equivalently, it can be expressed through the \textquotedblleft
q-characteristic function\textquotedblright\ (\ref{qGeneratingFunction})
leading to Eq. (\ref{FR_Generating}). Here, we search which kind of
distributions may satisfy these relations in an exact way for any value of $%
x.$

Gaussian distributions always satisfy the standard symmetry corresponding to 
$\lim q\rightarrow 1.$ Normal distributions emerge naturally when
formulating a central limit theorem, as solutions of linear Fokker-Planck
equations, or by maximizing Gibbs entropy under a second moment constraint.
Similarly, q-Gaussian distributions \cite{TsallisBook} are related to a
generalized central limit theorem \cite{QCentralLimit}, are solutions of a
kind of non-linear Fokker-Planck equations \cite{QFokker}, and maximize
nonextensive Tsallis entropy under a generalized second moment constraint 
\cite{qConstraints}. They read%
\begin{equation}
p(x)=\frac{\sqrt{\beta }}{\mathcal{N}_{q}}\exp _{q}[-\beta (x-x_{0})^{2}],
\label{qGaussiana}
\end{equation}%
where $\beta ^{-1}$ measures the width of the distribution and $\mathcal{N}%
_{q}$ is a normalization factor such that $\int_{-\infty }^{+\infty
}p(x)dx=1,$ $(q<3).$ For $1<q<3,$ $p(x)$ is characterized by power-law
tails, $p(x)\simeq 1/(x-x_{0})^{2/(q-1)},$ which is the case of interest in
this paper. Hence, the index $q$ determines the exponent of the power-law
tails. The first moment of $p(x)$\ is finite for $1<q<2,$ while the second
one for $1<q<5/3.$ In the domain $1<q<3,$ the normalization constant reads 
\cite{TsallisBook}%
\begin{equation}
\mathcal{N}_{q}=\sqrt{\frac{\pi }{q-1}}\ \frac{\Gamma (\frac{3-q}{2(q-1)})}{%
\Gamma (\frac{1}{q-1})},\ \ \ \ \ \ 1<q<3,  \label{Nq}
\end{equation}%
where $\Gamma (y)$ is the Gamma function.

It is immediate to prove that the generalized FR (\ref{FRLog}) [or Eq. (\ref%
{Prelation1})] is satisfied, for any value of $x,$ by the q-Gaussian
distribution (\ref{qGaussiana}) with%
\begin{equation}
\zeta =4x_{0}\frac{\beta ^{\frac{3-q}{2}}}{\mathcal{N}_{q}^{1-q}},\ \ \ \ \
\ \ \ \ \ \ \zeta _{ph}=4x_{0}\beta \Big{(}\frac{3-q}{2}\Big{)}.
\label{Zita}
\end{equation}%
In the second expression we used the integral%
\begin{equation}
\int_{-\infty }^{+\infty }[p(x)]^{q}dx=\Big{(}\frac{3-q}{2}\Big{)}\Big{(}%
\frac{\mathcal{N}_{q}}{\sqrt{\beta }}\Big{)}^{1-q},  \label{NormaQGauss}
\end{equation}%
valid for the distribution (\ref{qGaussiana}). When $q\rightarrow 1,$ it
follows the standard Gaussian expression $\zeta =\zeta _{ph}=4x_{0}\beta .$
On the other hand, it is simple to realize that an \textit{arbitrary}
distribution $p(x)$ satisfy the symmetry (\ref{FRLog}) when $x$ is
restricted to the interval where it develops power-law tails, $p(x)\simeq
1/(x-x_{0})^{2/(q-1)}.$ Hence, even in presence of power-law tails, a linear
relation characterize the symmetry between positive and negative
fluctuations of equal magnitude.

Using that $\int_{-\infty }^{+\infty }\exp _{q}(-ax^{2})=\mathcal{N}_{q}/%
\sqrt{a},$ after some algebra the q-characteristic function (\ref%
{qGeneratingFunction}) associated to the q-Gaussian distribution (\ref%
{qGaussiana}) reads%
\begin{equation}
Z_{q}(\lambda )=\exp _{\tilde{q}}\Big{\{}\!\Big{(}\frac{3-q}{2}\Big{)}\!%
\Big{(}\frac{\mathcal{N}_{q}}{\sqrt{\beta }}\Big{)}^{1-q}\Big{[}\frac{%
\lambda ^{2}\mathcal{N}_{q}^{1-q}}{4\beta ^{\frac{3-q}{2}}}-\lambda x_{0}%
\Big{]}\!\Big{\}},  \label{ZQGauss}
\end{equation}%
where the index of the exponential reads%
\begin{equation}
\tilde{q}=\frac{1+q}{3-q}.  \label{qTilde_qGauss}
\end{equation}%
Consistently, the function (\ref{ZQGauss}) satisfies the symmetry (\ref%
{FR_Generating}) with the constant $\zeta $ given by Eq. (\ref{Zita}). On
the other hand, $Z_{q}^{ph}(\lambda )$ reads%
\begin{equation}
Z_{q}^{ph}(\lambda _{ph})=\exp _{\tilde{q}}\Big{[}\frac{\lambda _{ph}^{2}}{%
4\beta (\frac{3-q}{2})}-\lambda _{ph}x_{0}\Big{]}.
\end{equation}

\subsection{Fluctuation relations for time-scaled variables}

While in the previous proposal we did not include time as an explicit
parameter, its generalization to time dependent variables is immediate, $%
x_{st}\rightarrow x_{st}(t),$ $p(x)\rightarrow p(x,t).$ On the other hand,
when studying nonequilibrium systems it is common to define the variable of
interest as a time-scaled one,%
\begin{equation}
\mu _{st}(t)\asymp \frac{1}{t^{\kappa }}\int_{0}^{t}dtv_{st}(t)\asymp \frac{%
x_{st}(t)}{t^{\kappa }},  \label{rate}
\end{equation}%
where $(d/dt)x_{st}(t)=v_{st}(t).$ Usually, the time scaling is proportional
to the elapsed time, $\kappa =1.$ Hence, $\mu _{st}(t)$ can be read as a
\textquotedblleft time-average\textquotedblright\ velocity. Here, we adopt a
more general point of view by considering arbitrary values of the exponent $%
\kappa >0.$ In general, the definition (\ref{rate}) makes sense in an
asymptotic time regime. From now on, we use the symbol $\asymp $ for
denoting an equality valid in a long time regime \cite{touchette}. Clearly,
the conditions that guarantees the achievement of this regime depend on each
particular system.

For variables such as $\mu _{st}(t)$ one can also define a generalized FR.
While its structure is very similar to the previous case, it is worthwhile
to write it explicitly. The probability distribution $p(\mu )$\ of $\mu
_{st}(t)$ follows from the change of variable $p(\mu )d\mu =p(x)dx.$ Taking
into account that $p(x)$ satisfies Eq. (\ref{FRLog}), a natural extension of
the generalized FR is 
\begin{equation}
\frac{1}{t^{\eta }}\{\ln _{q}[p(\mu )]-\ln _{q}[p(-\mu )]\}\asymp \alpha \mu
,  \label{FRProbaLong}
\end{equation}%
or equivalently%
\begin{equation}
p(-\mu )\asymp p(\mu )\otimes _{q}e_{q}^{-\alpha t^{\eta }\mu },\ \ \ \ \ \
p(\mu )\asymp p(-\mu )\otimes _{q}e_{q}^{\alpha \mu t^{\eta }}.
\label{qFRProbaMU}
\end{equation}%
The constant $\alpha $ as well as the exponent $\eta >0$ depend on each
particular problem. Similarly, the q-characteristic function of $\mu
_{st}(t) $ is defined as 
\begin{equation}
Z_{q}(\lambda )=\int_{-\infty }^{\infty }d\mu \lbrack p(\mu )\otimes
_{q}e_{q}^{-\lambda t^{\eta }\mu }].  \label{qCharacteristic}
\end{equation}%
From Eq. (\ref{qFRProbaMU}), it satisfies the symmetry%
\begin{equation}
Z_{q}(\lambda )\asymp Z_{q}(-\lambda +\alpha ),\ \ \ \ \ \ \ \ \ \ \ \ \ \
Z_{q}(\alpha )\asymp 1.  \label{FRCaracLong}
\end{equation}%
Notice that in order to simplify the notation we do not explicit either that 
$Z_{q}$ corresponds to the variable $\mu _{st}(t)$ or its dependence on time.

For the normalized scheme, Eq. (\ref{zitaPhys}), we define the equivalent FR%
\begin{equation}
\frac{1}{t^{\delta }}\{\ln _{q}[p(\mu )]-\ln _{q}[p(-\mu )]\}\asymp \frac{%
\alpha _{ph}}{\int_{-\infty }^{\infty }d\mu \lbrack p(\mu )]^{q}}\mu .
\label{BiassedFR}
\end{equation}%
In this case, the normalization time factor is defined with a different
exponent, $\delta >0.$ By introducing the characteristic function $%
Z_{q}^{ph}(\lambda _{ph})\equiv Z_{q}(\lambda (\lambda _{ph})),$ where%
\begin{equation}
\lambda =\frac{\lambda _{ph}}{\frac{1}{t^{\delta -\eta }}\int_{-\infty
}^{\infty }d\mu \lbrack p(\mu )]^{q}},  \label{lambdaMU}
\end{equation}%
it follows the equivalent symmetry%
\begin{equation}
Z_{q}^{ph}(\lambda _{ph})\asymp Z_{q}^{ph}(-\lambda _{ph}+\alpha _{ph}).
\end{equation}

When the probability distribution $p(\mu )$ has an asymptotic $(t\rightarrow
\infty )$ exponential structure $(q\rightarrow 1),$ the (standard) FR can be
analyzed through a large deviation theory \cite{sornette,touchette}. The
LDFs, that is, the factors that scales the time dependence of the
probability distributions and the characteristic functions, are related by a
Legendre-Fenchel transform. Furthermore, the relation (\ref{FRProbaLong}) $%
(q\rightarrow 1),$ or equivalently (\ref{BiassedFR}), implies some
symmetries for the LDFs \cite{spohn}. In the Appendixes, by establishing a
generalized large deviation-like principle, we demonstrate that these
results can be generalized to the present context $(q\neq 1).$ Furthermore,
general relations linking the unnormalized [Eq. (\ref{FRProbaLong})] and
normalized schemes [Eq. (\ref{BiassedFR})] are formulated [Eqs. (\ref{Neta})
and (\ref{RelacionALFAS})].

\section{Work performed on a particle immersed in a complex environment}

In the previous Section we have defined the alternative FR and in the
Appendixes we developed a related large deviation-like principle. Here, we
apply the proposal to a specific physical example. We study the fluctuations
of the work done on a dragged particle \cite{mazonka,zon} immersed in a
complex environment able to induce power-law tails in the particle
statistics \cite{cohenRapid,BeckCohenFT}. While this model has been studied
previously, the following analysis show the conceptual and physical
relevance of proposing an alternative FR consistent with a LDP.

The position $x(t)$ and velocity $v(t)$ of the particle obey the equations 
\begin{subequations}
\begin{eqnarray}
\frac{dx(t)}{dt} &=&v(t), \\
m\frac{dv(t)}{dt} &=&-\gamma v(t)-k[x(t)-x^{\ast }(t)]+\xi (t).
\end{eqnarray}%
The contribution $-\gamma v(t)$ gives the damping force. The term $%
-k[x(t)-x^{\ast }(t)]$\ is the force produced by an harmonic potential where
the position of its minimum is given by the arbitrary function $x^{\ast
}(t). $ Finally, the environment influence is introduced through the noise $%
\xi (t).$

In the overdamped regime, $mk\ll \gamma ^{2},$ the position evolution can be
approximated by the stochastic equation 
\end{subequations}
\begin{equation}
\gamma \frac{dx(t)}{dt}=-k[x(t)-x^{\ast }(t)]+\xi (t).  \label{overdamped}
\end{equation}%
The work performed on the particle by the harmonic drag force during a time $%
\tau $ is \cite{zon,cohenRapid}%
\begin{equation}
W_{\tau }=-k\int_{0}^{\tau }[x(t)-x^{\ast }(t)]v^{\ast }(t)dt,
\label{WorkEstocastico}
\end{equation}%
where $v^{\ast }(t)\equiv (d/dt)x^{\ast }(t).$

Independently of the environment model, the noise probability distribution
is symmetric around the origin. Hence, when it exists, the average noise
intensity is null, $\left\langle \xi (t)\right\rangle =0.$ By assuming valid
the same property for the initial particle position, $\left\langle
x(0)\right\rangle =0,$ we introduce the function%
\begin{equation}
M_{\tau }=k\int_{0}^{\tau }dtv^{\ast }(t)\int_{0}^{t}dt^{\prime
}e^{-(t-t^{\prime })/\tau _{0}}v^{\ast }(t^{\prime }),  \label{Mtau}
\end{equation}%
where the characteristic time is%
\begin{equation}
\tau _{0}\equiv \frac{\gamma }{k}.
\end{equation}%
If the distribution of the noise admits to define a first moment, it is
simple to realize that $M_{\tau }$ gives the average performed work, $%
M_{\tau }=\left\langle W_{\tau }\right\rangle .$ In general, it only defines
the most probable value of $W_{\tau }.$

For the Langevin dynamics (\ref{overdamped}), transient behaviors develops
when $\tau \lesssim \tau _{0}.$ Therefore, the long time regime is achieved
when $\tau \gg \tau _{0}.$ Under this condition, the exponential factor in
Eq. (\ref{Mtau}) can be approximated by a delta-Dirac function, $%
e^{-(t-t^{\prime })/\tau _{0}}\asymp \tau _{0}\delta (t-t^{\prime }).$
Hence, we get%
\begin{equation}
M_{\tau }\asymp \gamma \int_{0}^{\tau }dt[v^{\ast }(t)]^{2},
\label{WAverageLongTime}
\end{equation}%
where we have used that $k\tau _{0}=\gamma .$

The object of interest is the probability density $P_{\tau }(W)$ of
performing a work $W$ up to time $\tau .$ We also study the asymptotic
statistic of the dimensionless stochastic variable%
\begin{equation}
w_{\tau }\equiv \frac{W_{\tau }}{M_{\tau }},  \label{wChico}
\end{equation}%
whose probability density is denoted as $p_{\tau }(w).$ At any time, the
most probable value of $w_{\tau }$ is one. Equivalently, when it exists, $%
\left\langle w_{\tau }\right\rangle =1.$ Notice that $W_{\tau }$ and $%
w_{\tau }$ correspond respectively to the variables $x_{st}$ and $\mu _{st}$
of the previous section.

\subsection{Standard model}

In order to clarify the next results, here we briefly review the standard
case of a thermal environment at temperature $T.$ Therefore, $\xi (t)$ is a
Gaussian noise whose correlation, consistently with a
fluctuation-dissipation theorem, is $\left\langle \xi (t)\xi
(s)\right\rangle =2\beta ^{-1}\gamma \delta (t-s),$ where $\beta
^{-1}=k_{B}T.$ As demonstrated in Ref. \cite{zon}, for thermalized initial
conditions, $P_{\tau }(W)$ is a Gaussian distribution%
\begin{equation}
P_{\tau }(W)=\frac{1}{\sqrt{2\pi V_{\tau }}}\exp \Big{[}-\frac{(W-M_{\tau
})^{2}}{2V_{\tau }}\Big{]},  \label{PWGauss}
\end{equation}%
where the time dependent width $V_{\tau }$ is%
\begin{equation}
V_{\tau }=2\beta ^{-1}M_{\tau }.  \label{FDRGauss}
\end{equation}%
$P_{\tau }(W)$ satisfies the standard FR symmetry%
\begin{equation}
\ln [P_{\tau }(W)]-\ln [P_{\tau }(-W)]=\beta W.  \label{FRWorkStandard}
\end{equation}%
The proportionality with $\beta $ arises owe to the relation (\ref{FDRGauss}%
).

After a simple change of variables, the probability density of (\ref{wChico}%
) is%
\begin{equation}
p_{\tau }(w)=\frac{1}{\sqrt{2\pi \tilde{V}_{\tau }}}\exp \Big{[}-\frac{%
(w-w_{0})^{2}}{2\tilde{V}_{\tau }}\Big{]},
\end{equation}%
where%
\begin{equation}
w_{0}=1,\ \ \ \ \ \ \ \ \tilde{V}_{\tau }=\frac{V_{\tau }}{M_{\tau }^{2}}=%
\frac{2}{\beta M_{\tau }}.
\end{equation}%
Hence, it follows the symmetry%
\begin{equation}
\frac{1}{\tau ^{\delta }}\{\ln [p_{\tau }(w)]-\ln [p_{\tau }(-w)]\}=\frac{%
\beta M_{\tau }}{\tau ^{\delta }}w\asymp \alpha _{ph}w.
\label{FRStandardwChica}
\end{equation}%
The exponent $\delta =\eta $ is chosen in such a way that in the asymptotic
regime the contribution $M_{\tau }/\tau ^{\delta }$ does not depends on
time. We assume an accelerated potential movement%
\begin{equation}
x^{\ast }(t)=\frac{v_{\ast }t^{1+a_{c}}}{(1+a_{c})},\ \ \ \ \ \ \ \ \
v^{\ast }(t)=v_{\ast }t^{a_{c}},  \label{aceleracion}
\end{equation}%
where $v_{\ast }$ is an appropriate constant and $a_{c}\geq 0.$ The usual
assumption of constant velocity, $v^{\ast }(t)=v_{\ast },$ is recovered with 
$a_{c}=0.$ The average work (\ref{WAverageLongTime})\ behaves as%
\begin{equation}
M_{\tau }\asymp \frac{\gamma v_{\ast }^{2}}{1+2a_{c}}\tau ^{1+2a_{c}}.
\end{equation}%
Therefore, the FR (\ref{FRStandardwChica})\ is defined with%
\begin{equation}
\delta =1+2a_{c},\ \ \ \ \ \ \ \ \alpha _{ph}=\frac{1}{1+2a_{c}}\beta \gamma
v_{\ast }^{2}.
\end{equation}%
Notice that in general $\delta \neq 1.$ A similar situation emerges in
presence of normal long time correlated fluctuations \cite{klages,harris}.
Here, this dependence arises from the acceleration of the harmonic
potential, Eq. (\ref{aceleracion}). In fact, $\delta =1$ when the velocity
is constant, $a_{c}=0.$ On the other hand, owe to the chosen normalization,
Eq. (\ref{wChico}), $\alpha _{ph}$ is not only proportional to $\beta $
(factor $\gamma v_{\ast }^{2}).$

\subsection{Superstatistical model}

Superstatistics \cite{beckQ,superstatistics} consists of superpositions of
different statistics for driven nonequilibrium system with spatiotemporal
inhomogeneities of an intensive parameter, such as for example the inverse
temperature $\beta $ in the previous example. We assume that the time scale
on which $\beta $ fluctuates is much larger than the typical fluctuations of 
$W_{\tau },$ that is, $\tau _{0}=\gamma /k.$ Hence, the work distribution
can be written as $P_{\tau }(W)\asymp \int d\beta f(\beta )P_{\tau
}^{G}(W,\beta ),$ where $P_{\tau }^{G}(W,\beta )$ is the probability density
of the normal environment case, Eq. (\ref{PWGauss}). This limit was analyzed
in Ref. \cite{BeckCohenFT} without resorting on a LDP. A very complex
probability relation arises, having not any clear dependence with the
average environment temperature. These drawbacks are surpassed with the
present approach.

As in Ref. \cite{beckQ}, the temperature fluctuations are described by a
Gamma distribution%
\begin{equation}
f(\beta )=\frac{1}{\Gamma (n/2)}\Big{(}\frac{n}{2\beta _{0}}\Big{)}%
^{n/2}\beta ^{\frac{n}{2}-1}\exp \Big{(}-\frac{n\beta }{2\beta _{0}}\Big{)},
\label{DistriTemper}
\end{equation}%
where $\beta _{0}$ corresponds to the average inverse temperature, $\beta
_{0}=\int d\beta f(\beta )\beta ,$ and $n$ is a positive constant. After
averaging over this distribution, it follows the q-Gaussian distribution%
\begin{equation}
P_{\tau }(W)\asymp \frac{1}{\sqrt{2\mathcal{N}_{q}^{2}V_{\tau }}}\exp _{q}%
\Big{[}-\frac{(W-M_{\tau })^{2}}{2V_{\tau }}\Big{]}.  \label{qGaussWSuper}
\end{equation}%
Here, the width function $V_{\tau }$ is given by%
\begin{equation}
V_{\tau }=2\beta _{q}^{-1}M_{\tau }.
\end{equation}%
Notice the similitude with Eq. (\ref{FDRGauss}). Here, $M_{\tau }$ also
follows from Eq. (\ref{WAverageLongTime}). On the other hand, the
characteristic parameters are \cite{beckQ}%
\begin{equation}
q=1+\frac{2}{n+1},\ \ \ \ \ \ \ \ \ \beta _{q}=\frac{2}{3-q}\beta _{0}.
\label{AverageTemperature}
\end{equation}

It is difficult to extract some physical information by analyzing the
distribution (\ref{qGaussWSuper}) through a standard FR symmetry \cite%
{BeckCohenFT}, Eq. (\ref{FRWorkStandard}). In fact, given that $P_{\tau }(W)$
is a q-Gaussian distribution, it satisfies the generalized FR (\ref{FRLog}).
The parameter $\zeta ,$ from Eqs. (\ref{Zita}) and (\ref{qGaussWSuper}) can
be written as%
\begin{equation}
\zeta =\Big{[}2\mathcal{N}_{q}\sqrt{M_{\tau }}\Big{]}^{q-1}\beta _{q}^{\frac{%
3-q}{2}},\ \ \ \ \ \ \ \ \ \zeta _{ph}=\beta _{0}.  \label{ZitaSuper}
\end{equation}%
While the unnormalized FR is defined by a linear dependence on $W,$ $\zeta $
depends on time. Hence, its physical content is unclear. Nevertheless, in
the normalized scheme, $\zeta _{ph}$ corresponds to the physical (average)
temperature $\beta _{0}.$ Explicitly, the work probability distribution
satisfies the FR symmetry%
\begin{equation}
\mathcal{Z}_{q}^{\tau }\{\ln _{q}[P_{\tau }(W)]-\ln _{q}[P_{\tau
}(-W)]\}\asymp \beta _{0}W,  \label{NuevaWFR}
\end{equation}%
where for notational convenience we have defined%
\begin{equation}
\mathcal{Z}_{q}^{\tau }\equiv \int_{-\infty }^{\infty }dW[P_{\tau }(W)]^{q}.
\end{equation}%
As in the standard case, in the present approach the environment (average)
temperature is the scaling parameter of the probabilities linear relation (%
\ref{NuevaWFR}). Furthermore, this novel and elegant result recovers the
standard FR (\ref{FRWorkStandard}) in the limit $q\rightarrow 1,$ that is,
in absence of temperature fluctuations, $f(\beta )=\delta (\beta -\beta
_{0}).$

The probability density of $w_{\tau },$ Eq. (\ref{wChico}), is%
\begin{equation}
p_{\tau }(w)\asymp \frac{1}{\sqrt{2\mathcal{N}_{q}^{2}\tilde{V}_{\tau }}}%
\exp _{q}\Big{[}-\frac{(w-w_{0})^{2}}{2\tilde{V}_{\tau }}\Big{]},
\label{pwNormalSuperstatistics}
\end{equation}%
where the coefficients read%
\begin{equation}
w_{0}=1,\ \ \ \ \ \ \ \ \tilde{V}_{\tau }=\frac{V_{\tau }}{M_{\tau }^{2}}=%
\frac{2}{\beta _{q}M_{\tau }}.
\end{equation}%
Hence, $p_{\tau }(w)$ satisfies the symmetry%
\begin{equation}
\frac{\mathit{z}_{q}^{\tau }}{\tau ^{\delta }}\{\ln _{q}[p_{\tau }(w)]-\ln
_{q}[p_{\tau }(-w)]\}\asymp \frac{\beta _{0}M_{\tau }}{\tau ^{\delta }}%
w\asymp \alpha _{ph}w,  \label{RectaNormalSupestatistical}
\end{equation}%
where the normalization reads%
\begin{equation}
\mathit{z}_{q}^{\tau }\equiv \int_{-\infty }^{\infty }dw[p_{\tau }(w)]^{q}.
\label{zchica}
\end{equation}%
This integral can be performed by using Eq. (\ref{NormaQGauss}). By assuming
the velocity dependence (\ref{aceleracion}), it follows%
\begin{equation}
\delta =1+2a_{c},\ \ \ \ \ \ \ \ \alpha _{ph}=\frac{1}{1+2a_{c}}\beta
_{0}\gamma v_{\ast }^{2}.  \label{DeltaAlfa}
\end{equation}

The previous analysis demonstrate that all features of a standard FR remain
valid for the present example if one use the generalized FR. This result is
explicitly shown in Fig. 1.%
%figura%figura%figura%figura%figurav%figura%figura%figura%figura%figura%figura%figura%figura%figura%figurav%figura%figura%figura%figura%figura
%figura%figura%figura%figura%figurav%figura%figura%figura%figura%figura%figura%figura%figura%figura%figurav%figura%figura%figura%figura%figura
\begin{figure}[tbp]
\includegraphics[bb=30 335 600 1120,angle=0,width=7.5cm]{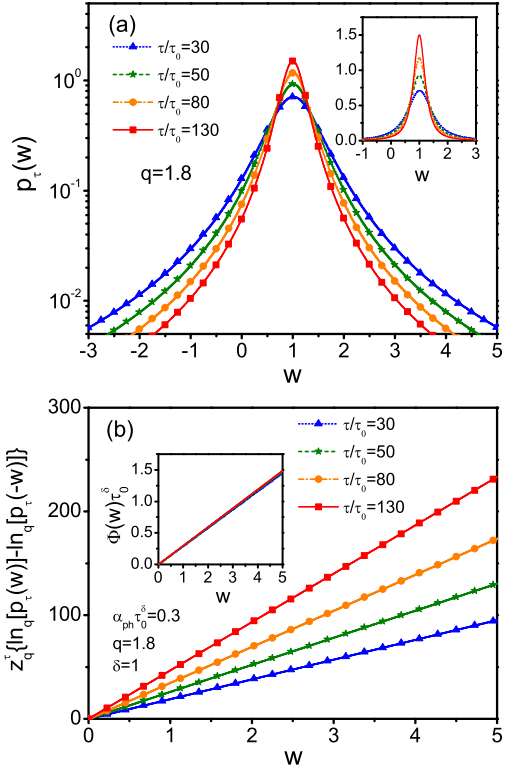}
\caption{(Color Online) (a) Probability distribution $p_{\protect\tau }(w)$
for the superstatistical model, Eq. (\protect\ref{pwNormalSuperstatistics}),
for different times $\protect\tau .$ The inset details the peak region. (b)
Plot of $\mathit{z}_{q}^{\protect\tau }\{\ln _{q}[p_{\protect\tau }(w)]-\ln
_{q}[p_{\protect\tau }(-w)]\}$ for each time $\protect\tau .$ The inset
shows the collapse of these curves to a single line when plotting the
normalized FR symmetry $\Phi (w)\equiv \mathit{z}_{q}^{\protect\tau }\{\ln
_{q}[p_{\protect\tau }(w)]-\ln _{q}[p_{\protect\tau }(-w)]\}/\protect\tau ^{%
\protect\delta },$ Eq. (\protect\ref{RectaNormalSupestatistical}). The
potential velocity is constant, $a_{c}=0,$\ Eq. (\protect\ref{aceleracion}).
Hence, $\protect\delta =1,$ Eq. (\protect\ref{DeltaAlfa}). In the natural
units of the problem (see text) the parameters are $v_{\ast }=\protect\gamma %
=k=1.$ The average temperature is $\protect\beta _{0}\protect\gamma v_{\ast
}^{2}/k=\protect\alpha _{ph}\protect\tau _{0}=0.3,$ and $q=1.8,$ [Eq.(%
\protect\ref{AverageTemperature})].}
\end{figure}
%figura%figura%figura%figura%figurav%figura%figura%figura%figura%figura%figura%figura%figura%figura%figurav%figura%figura%figura%figura%figura
%figura%figura%figura%figura%figurav%figura%figura%figura%figura%figura%figura%figura%figura%figura%figurav%figura%figura%figura%figura%figura

In Fig. 1(a) we plot the q-Gaussian distribution (\ref%
{pwNormalSuperstatistics}) for different times $\tau .$ In the inset we show
the peaks of the distributions. Their wide diminish with time. We assumed a
constant potential's velocity; $a_{c}=0$ in Eq. (\ref{aceleracion}). Hence, $%
\delta =1.$ By using the (natural) units of mass $m_{0}=\gamma ^{2}/k,$
distance $x_{0}=v_{\ast }\gamma /k,$ and time $\tau _{0}=\gamma /k,$ it
follows $\gamma =k=v_{\ast }=1.$ Therefore, the unique free parameters are $%
q $ [Eq. (\ref{AverageTemperature})] and the (dimensionless) noise
intensity, that is, the average temperature of the distribution (\ref%
{DistriTemper}), $\beta _{0}(\gamma v_{\ast })^{2}/k=\alpha _{ph}\tau _{0}.$

In Fig. 1(b) we plot the dependence with $w$ of $\mathit{z}_{q}^{\tau }\{\ln
_{q}[p_{\tau }(w)]-\ln _{q}[p_{\tau }(-w)]\}.$ The index $q$ is the same
that defines the q-Gaussian distribution of Fig. 1(a). For each time, a
linear behavior is evident. In the inset we show the collapse to a single
line when introducing the time normalization factor $(1/\tau ^{\delta }),$
Eq. (\ref{RectaNormalSupestatistical}).

In the unnormalized scheme [Eq. (\ref{FRProbaLong})], the FR reads%
\begin{equation}
\frac{1}{\tau ^{\eta }}\{\ln _{q}[p_{\tau }(w)]-\ln _{q}[p_{\tau
}(-w)]\}\asymp \frac{\lbrack \beta _{q}M_{\tau }]^{\frac{3-q}{2}}}{\tau
^{\eta }(2\mathcal{N}_{q})^{1-q}}\asymp \alpha w,
\end{equation}%
where the coefficients are%
\begin{equation}
\alpha =\frac{1}{(2\mathcal{N}_{q})^{1-q}}\Big{(}\frac{\beta _{q}\gamma
v_{\ast }^{2}}{1+2a_{c}}\Big{)}^{\frac{3-q}{2}},\ \ \ \ \ \ \ \ \eta =\delta %
\Big{(}\frac{3-q}{2}\Big{)}.
\end{equation}%
Here, the coefficient $\alpha $ does not have a clear physical meaning.
Nevertheless, as commented before, based on a large deviation-like theory
(Appendixes) it is possible to establish some general relations between $%
\alpha $ and $\alpha _{ph,}$\ as well as between $\eta $ and $\delta $ [Eqs.
(\ref{Neta}) and (\ref{RelacionALFAS})], which are satisfied in the present
case.

\subsection{L\'{e}vy noise model}

In Ref. \cite{cohenRapid} the noise $\xi (t)$ was taken as a symmetric
stable L\'{e}vy noise. Furthermore, different experimental setups where the
model may be explicitly measured were proposed. While the analysis presented
in that contribution is completely right, here we study the same problem by
using the generalized FR. We explicitly show that a FR can be established
only when the probabilities satisfy a LDP.

The noise is defined by its characteristic functional 
\begin{subequations}
\label{Functional}
\begin{eqnarray}
G_{\xi }[\lambda (t)] &=&\Big\langle\exp i\int_{0}^{\infty }dt\lambda (t)\xi
(t)\Big\rangle, \\
&=&\exp \Big{(}-b\int_{0}^{\infty }dt|\lambda (t)|^{\sigma }\Big{)},
\end{eqnarray}%
where $\lambda (t)$ is an arbitrary test function. The constant $b$ measures
the noise intensity and $0<\sigma <2.$ Due to the linearity of the
stochastic dynamics (\ref{overdamped}), the work (\ref{WorkEstocastico}) is
also a stable variable with the same index $\sigma .$ It characteristic
function, $G_{W}(\lambda )=\int_{-\infty }^{+\infty }dWP_{\tau }(W)\exp
[i\lambda W],$ then reads 
\end{subequations}
\begin{subequations}
\label{WCharacLevy}
\begin{eqnarray}
G_{W}(\lambda ) &=&\langle \exp [i\lambda W_{\tau }]\rangle , \\
&=&\exp (iM_{\tau }\lambda -B_{\tau }\left\vert \lambda \right\vert ^{\sigma
}),
\end{eqnarray}%
where $M_{\tau }$ is defined by Eq. (\ref{Mtau}) and $B_{\tau }$ can be
obtained after writing $G_{W}(\lambda )$ in terms of $G_{\xi }[\lambda (t)].$
For arbitrary velocities $v^{\ast }(t),$ we get 
\end{subequations}
\begin{equation}
B_{\tau }=b\Big{(}\frac{k}{\gamma }\Big{)}^{\sigma }\int_{0}^{\tau
}dt\left\vert \int_{t}^{\tau }dt^{\prime }e^{-(t^{\prime }-t)/\tau
_{0}}v^{\ast }(t^{\prime })\right\vert ^{\sigma }.
\end{equation}%
For $\tau \gg \tau _{0},$ $M_{\tau }$ can be calculated from Eq. (\ref%
{WAverageLongTime}), while $B_{\tau },$ after taking $e^{-(t^{\prime
}-t)/\tau _{0}}\asymp \tau _{0}\delta (t^{\prime }-t),$ can be approximated
as%
\begin{equation}
B_{\tau }\asymp b\int_{0}^{\tau }dt\left\vert v^{\ast }(t)\right\vert
^{\sigma }.  \label{VLevyAsymptotico}
\end{equation}

Eq. (\ref{WCharacLevy}) corresponds to the Fourier transform of a L\'{e}vy
probability distribution. As is well known \cite{levyNum}, for $|W-M_{\tau
}|\gtrsim B_{\tau }^{1/\sigma }$ it develops power-law tails,%
\begin{equation}
P_{\tau }(W)\approx \frac{c_{\sigma }B_{\tau }}{|W-M_{\tau }|^{1+\sigma }},
\label{ColasDeLevy}
\end{equation}%
where $c_{\sigma }=\pi ^{-1}\sigma \sin (\pi \sigma /2)\Gamma (\sigma ).$
Only when $\sigma =1,$ one gets a simple analytical expression valid for any
value of $W,$ $P_{\tau }(W)=(V_{\tau }/\pi )[(W-M_{\tau })^{2}+V_{\tau
}^{2}]^{-1}.$ It is expected that Eq. (\ref{ColasDeLevy}) satisfies the
normalized FR%
\begin{equation}
\mathcal{Z}_{q}^{\tau }\{\ln _{q}[P_{\tau }(W)]-\ln _{q}[P_{\tau
}(-W)]\}\propto \zeta _{ph}(\tau )W,  \label{LevyNormaQFR}
\end{equation}%
where the symbol $\propto $ denotes both an asymptotic time regime $(\tau
\gg \tau _{0})$ and $|W-M_{\tau }|\gtrsim B_{\tau }^{1/\sigma },$ that is,
values of $W$ in the power-law regime. The parameter $q$ and the function $%
\zeta _{ph}(\tau )$ can be found by mapping the approximation (\ref%
{ColasDeLevy}) with the power-law behavior of the q-Gaussian distribution (%
\ref{qGaussWSuper}). We get%
\begin{equation}
q=\frac{\sigma +3}{\sigma +1},\ \ \ \ \ \ \ \ \ V_{\tau }=c_{\sigma
}^{\prime }B_{\tau }^{\frac{2}{\sigma }}.  \label{mapaQSigma}
\end{equation}%
where $c_{\sigma }^{\prime }=\frac{1}{2}(\frac{2}{1+\sigma })^{\frac{%
1+\sigma }{\sigma }}(c_{\sigma }\mathcal{N}_{q})^{\frac{2}{\sigma }}.$
Notice that $5/3<q<3$ \cite{LevyAsQ}. With these relations at hand, the time
dependent function $\zeta (\tau )$ reads%
\begin{equation}
\zeta _{ph}(\tau )\approx c_{\sigma }^{\prime \prime }\frac{\gamma }{b^{%
\frac{2}{\sigma }}}\frac{1}{\tau ^{\frac{2}{\sigma }-1}}.
\label{ZitaLevyVanishing}
\end{equation}%
with $c_{\sigma }^{\prime \prime }=\frac{2}{c_{\sigma }^{\prime }}\frac{%
\sigma }{1+\sigma }\frac{(1+a_{c}\sigma )^{\frac{2}{\sigma }}}{1+2a_{c}}.$
In deriving this expression we assumed the general velocity dependence (\ref%
{aceleracion}). While in the power-law regime the L\'{e}vy distribution
satisfy the generalized FR (\ref{LevyNormaQFR}), the proportionality
constant [Eq. (\ref{ZitaSuper})] becomes time dependent, $\zeta
_{ph}\rightarrow \zeta _{ph}(\tau ).$ At long times, for any value of $%
\sigma \in (0,2),$ it vanishes. Hence, consistently with the results of Ref. 
\cite{cohenRapid}, one conclude that asymptotically positive and negative
fluctuations of the same magnitude have the same statistical weight. On the
other hand, by comparison with the fluctuation theorem (\ref{NuevaWFR}), it
follows that here it is not possible to associate a temperature to the
stochastic L\'{e}vy dynamics.

Other distinctive features of the problem can be characterized by analyzing
the statistics of the dimensionless work (\ref{wChico}). After a simple
changes of variables, from Eq. (\ref{WCharacLevy}), the Fourier transform of
its probability $p_{\tau }(w)$ reads 
\begin{subequations}
\label{LevyPw}
\begin{eqnarray}
G_{w}(\lambda ) &=&\langle \exp [i\lambda w_{\tau }]\rangle , \\
&=&\exp (i\lambda w_{0}-\tilde{B}_{\tau }\left\vert \lambda \right\vert
^{\sigma }),
\end{eqnarray}%
where the coefficients are 
\end{subequations}
\begin{equation}
w_{0}=1,\ \ \ \ \ \ \ \ \tilde{B}_{\tau }=\frac{B_{\tau }}{|M_{\tau
}|^{\sigma }}.
\end{equation}%
From Eqs. (\ref{WAverageLongTime}) and (\ref{VLevyAsymptotico}) we get the
asymptotic behavior%
\begin{equation}
\tilde{B}_{\tau }\asymp \frac{b}{\gamma ^{\sigma }}\frac{\int_{0}^{\tau
}dt\left\vert v^{\ast }(t)\right\vert ^{\sigma }}{\left\vert \int_{0}^{\tau
}dt[v^{\ast }(t)]^{2}\right\vert ^{\sigma }}.
\end{equation}%
For the velocity dependence (\ref{aceleracion}), it follows%
\begin{equation}
\tilde{B}_{\tau }\asymp d_{\sigma }\frac{b}{(\gamma v_{\ast })^{\sigma }}%
\frac{1}{\tau ^{\sigma (1+a_{c})-1}},\ \ \ \ \ \ \ v^{\ast }(t)=v_{\ast
}t^{a_{c}}.  \label{BTildeLevy}
\end{equation}%
where $d_{\sigma }=\frac{(1+2a_{c}\sigma )^{\sigma }}{1+a_{c}\sigma }.$
Therefore, when $\sigma >1/(1+a_{c})$ the width of the distribution
diminishes, while for $\sigma <1/(1+a_{c})$ it increases with time. The
former behavior is consistent with a LDP [see Eq. (\ref{AsymptoticP})].
Therefore, it should be possible to establishes a (generalized) FR symmetry
for $p_{\tau }(w).$ In the second case, the typical fluctuations of the
scaled work $w_{\tau }$ increase in size at higher times. This anomalous
behavior is inconsistent with both a LDP and the law of large numbers \cite%
{cohenRapid}. Therefore, we expect that $p_{\tau }(w)$ does not fulfill any
FR in this case. On the other hand, by analyzing $p_{\tau }(w)$ through a
standard FR, the transition $\sigma \gtrless 1$ $(a_{c}=0)$ leads to the
different characteristic behaviors found in Ref. \cite{cohenRapid}.%
%figura%figura%figura%figura%figurav%figura%figura%figura%figura%figura%figura%figura%figura%figura%figurav%figura%figura%figura%figura%figura
%figura%figura%figura%figura%figurav%figura%figura%figura%figura%figura%figura%figura%figura%figura%figurav%figura%figura%figura%figura%figura
\begin{figure}[tbp]
\includegraphics[bb=30 335 600
1120,angle=0,width=7.5cm]{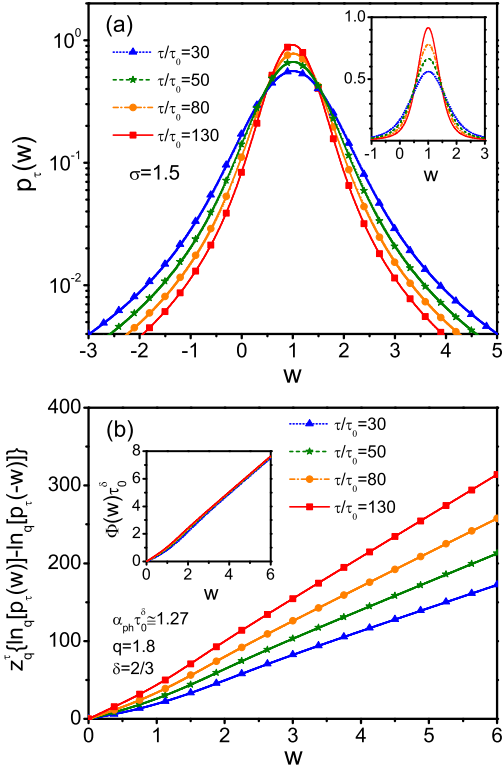}
\caption{(Color Online) (a) Levy probability distribution $p_{\protect\tau %
}(w)$ defined by Fourier transform (\protect\ref{LevyPw}) with $\protect%
\sigma =1.5.$ The inset details the peak region. (b) Plot of $\mathit{z}%
_{q}^{\protect\tau }\{\ln _{q}[p_{\protect\tau }(w)]-\ln _{q}[p_{\protect%
\tau }(-w)]\}$ for each time $\protect\tau .$ The inset shows the collapse
of these curves when plotting the normalized FR symmetry $\Phi (w)\equiv 
\mathit{z}_{q}^{\protect\tau }\{\ln _{q}[p_{\protect\tau }(w)]-\ln _{q}[p_{%
\protect\tau }(-w)]\}/\protect\tau ^{\protect\delta },$ Eq. (\protect\ref%
{FRGeneLevy}). The potential velocity is constant, $a_{c}=0,$\ Eq. (\protect
\ref{aceleracion}). Hence, $\protect\delta =2/3,$ Eq. (\protect\ref%
{DeltaLevyNormalizada}). In natural units (see text) the parameters are $%
v_{\ast }=\protect\gamma =k=1.$ The noise intensity reads $b(\protect\gamma %
v_{\ast })^{2}=2.$}
\end{figure}
%figura%figura%figura%figura%figurav%figura%figura%figura%figura%figura%figura%figura%figura%figura%figurav%figura%figura%figura%figura%figura
%figura%figura%figura%figura%figurav%figura%figura%figura%figura%figura%figura%figura%figura%figura%figurav%figura%figura%figura%figura%figura
%figura%figura%figura%figura%figurav%figura%figura%figura%figura%figura%figura%figura%figura%figura%figurav%figura%figura%figura%figura%figura
%figura%figura%figura%figura%figurav%figura%figura%figura%figura%figura%figura%figura%figura%figura%figurav%figura%figura%figura%figura%figura
\begin{figure}[tbp]
\includegraphics[bb=30 335 600
1120,angle=0,width=7.5cm]{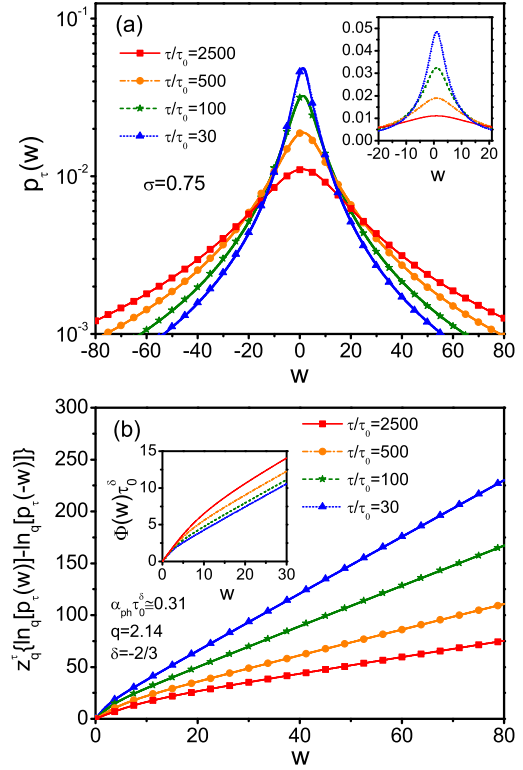}
\caption{(Color Online) (a) Levy probability distribution $p_{\protect\tau %
}(w)$ defined by the Fourier transform (\protect\ref{LevyPw}) with $\protect%
\sigma =0.75.$ The inset details the peak region. (b) Plot of $\mathit{z}%
_{q}^{\protect\tau }\{\ln _{q}[p_{\protect\tau }(w)]-\ln _{q}[p_{\protect%
\tau }(-w)]\}$ for each time $\protect\tau .$ The inset corresponds to $\Phi
(w)\equiv \mathit{z}_{q}^{\protect\tau }\{\ln _{q}[p_{\protect\tau }(w)]-\ln
_{q}[p_{\protect\tau }(-w)]\}/\protect\tau ^{\protect\delta },$ Eq. (\protect
\ref{FRGeneLevy}). The potential velocity is constant, $a_{c}=0,$\ Eq. (%
\protect\ref{aceleracion}). Hence, $\protect\delta =-2/3<0,$ Eq. (\protect
\ref{DeltaLevyNormalizada}). In natural units (see text) the parameters are $%
v_{\ast }=\protect\gamma =k=1.$ The noise intensity reads $b(\protect\gamma %
v_{\ast })^{2}=2.$}
\end{figure}
%figura%figura%figura%figura%figurav%figura%figura%figura%figura%figura%figura%figura%figura%figura%figurav%figura%figura%figura%figura%figura
%figura%figura%figura%figura%figurav%figura%figura%figura%figura%figura%figura%figura%figura%figura%figurav%figura%figura%figura%figura%figura

For $|w-w_{0}|\gtrsim \tilde{B}_{\tau }^{1/\sigma },$ the probability
distribution $p_{\tau }(w)$ behaves as $p_{\tau }(w)\approx c_{\sigma }%
\tilde{B}_{\tau }/|w-w_{0}|^{1+\sigma }$ [see Eq. (\ref{ColasDeLevy})]. In
that regime, it satisfies the relation%
\begin{equation}
\frac{\mathit{z}_{q}^{\tau }}{\tau ^{\delta }}\{\ln _{q}[p_{\tau }(w)]-\ln
_{q}[p_{\tau }(-w)]\}\propto \alpha _{ph}w,  \label{FRGeneLevy}
\end{equation}%
where the coefficients read%
\begin{equation}
\delta =2(1+a_{c}-\sigma ^{-1}),\ \ \ \ \ \ \ \ \alpha _{ph}\approx
d_{\sigma }^{\prime }b^{-\frac{2}{\sigma }}(\gamma v_{\ast })^{2},
\label{DeltaLevyNormalizada}
\end{equation}%
with $d_{\sigma }^{\prime }=\frac{2\sigma }{1+\sigma }(c_{\sigma }^{\prime
}d_{\sigma }^{2/\sigma })^{-1},$ and $q$ is defined by Eq. (\ref{mapaQSigma}%
), that is, $q=\frac{\sigma +3}{\sigma +1}.$ On the other hand, the
expression for $\alpha _{ph}$ is only approximated because it is based on a
mapping with a q-Gaussian distribution. In general, the value of $\mathit{z}%
_{q}^{\tau },$ Eq. (\ref{zchica}), differs from that corresponding to a L%
\'{e}vy distribution.

Eq. (\ref{FRGeneLevy}) is valid for any value of $\sigma \in (0,2)$ if $w$
pertains to the power-law domain. Equivalently, it does not applies for $%
|w-w_{0}|\lesssim \tilde{B}_{\tau }^{1/\sigma }.$ Therefore, when $\tilde{B}%
_{\tau }$ decrease (increase) in time the FR symmetry is valid (not valid)
for almost any value of $w.$ In fact, the transition in the behavior of the
characteristic width $\tilde{B}_{\tau },$ Eq. (\ref{BTildeLevy}), also
determines when a LDP applies $(\delta >0)$ or not $(\delta <0).$ From Eq. (%
\ref{DeltaLevyNormalizada}) it follows $\delta \gtrless 0$ if $\sigma
\gtrless 1/(1+a_{c}).$ These properties are explicitly shown in the next
figures.

In Fig. 2(a) we plot the (exact numeric) L\'{e}vy distribution $p_{\tau }(w)$
\cite{levyNum} obtained from its Fourier transform (\ref{LevyPw}). We
assumed a constant potential's velocity, $a_{c}=0$ in Eq. (\ref{aceleracion}%
), and $\sigma =1.5.$\ Hence, $q=1.8$ [Eq. (\ref{mapaQSigma})], and $\delta
=2/3>0$ [Eq. (\ref{DeltaLevyNormalizada})]. Consistently, the peaks around $%
w=1$ diminish their wide for higher times. By using natural units, the
unique free parameter is the (dimensionless) noise intensity [Eq. (\ref%
{Functional})], $b\tau _{0}(\gamma ^{2}v_{\ast }/k)^{-\sigma }.$

In Fig. 2(b) we plot the dependence with $w$ of $\mathit{z}_{q}^{\tau }\{\ln
_{q}[p_{\tau }(w)]-\ln _{q}[p_{\tau }(-w)]\}$ for different times $\tau .$\
Small deviations with respect to a linear behavior are observed around $%
|w_{0}-\tilde{B}_{\tau }^{1/\sigma }|<w<|w_{0}+\tilde{B}_{\tau }^{1/\sigma
}|.$ Their magnitude diminish with time. In the inset we show the collapse
to a single curve when introducing the time normalization factor $(1/\tau
^{\delta }),$ Eq. (\ref{FRGeneLevy}). We checked that the same property is
valid for higher times, indicating the consistence between the generalized
FR and its associated LDP. The value of $\alpha _{ph}\tau _{0}^{\delta }$
was estimated from the slope of the collapsed curves (inset). The
theoretical estimation, Eq. (\ref{DeltaLevyNormalizada}), gives $\alpha
_{ph}\tau _{0}^{\delta }=1.85.$

In Fig. 3(a) we plot the L\'{e}vy distribution $p_{\tau }(w)$ for $\sigma
=0.75,$ and $a_{c}=0.$ Hence, $q=15/7\simeq 2.14$ [Eq. (\ref{mapaQSigma})],
and $\delta =-2/3<0$ [Eq. (\ref{DeltaLevyNormalizada})]. A negative $\delta $
implies that the wide of the peaks around $w=1$ grows with time. The linear
behavior of $\mathit{z}_{q}^{\tau }\{\ln _{q}[p_{\tau }(w)]-\ln _{q}[p_{\tau
}(-w)]\}$ with $w$ is only valid for $w\gtrsim w_{0}+\tilde{B}_{\tau
}^{1/\sigma },$ where $\tilde{B}_{\tau }$ increases in time. Added to this
failure, after introducing the normalization factor $(1/\tau ^{\delta }),$
the curves does not collapse into a single curve (inset). These properties
are parallel to the inapplicability of a LDP. On the other hand, the
theoretical estimation, Eq. (\ref{DeltaLevyNormalizada}), gives $\alpha
_{ph}\tau _{0}^{\delta }=0.21.$

\section{Summary and Conclusions}

We have introduced an alternative definition of FR symmetry that satisfies
two conditions. In the regime where the probability of interest develops
power-law tails the symmetry is expressed through a linear behavior.
Furthermore, the generalized symmetry has associated a large deviation-like
theory.

The FR symmetry can be written as a difference between the generalized
q-logarithm of the probability distributions for positive and negative
fluctuations, Eq. (\ref{FRLog}). The parameter $q$ depends on the exponent
of the power-law tails. In terms of a generalized characteristic function,
Eq. (\ref{qGeneratingFunction}), the proposed FR can be expressed as in the
standard case, Eq. (\ref{FR_Generating}). Similar relations, Eqs. (\ref%
{FRProbaLong}) and (\ref{BiassedFR}), were formulated for time-scaled
variables. Based on large deviation-like principle, in the Appendixes we
showed that a set of LDFs can be consistently defined for both the
probability distribution and its associated characteristic function. The
standard Legendre structure connecting them remains valid even in presence
of self-similar power-law distributions, Eqs. (\ref{PhysTeTa}) and (\ref%
{PhysFi}). Therefore, the generalized FR can be expressed as in the standard
case when written in term of the LDFs, Eqs. (\ref{FR_PhysFI}) and (\ref%
{Fr_PhysTeta}).

The general formalism was applied for characterizing the fluctuations of the
work performed on a dragged particle immersed in a complex environment. When
the power-law nature of the dynamics is induced by (spaciotemporal)
temperature fluctuations, the work statistics is given by a q-Gaussian
distribution. The FR symmetry is scaled by the environment average
temperature. This novel fluctuation theorem [Eq. (\ref{NuevaWFR})] may in
principle be confirmed in different experimental setups \cite{beckQ}.

By analyzing the case in which the environment is represented by an external
L\'{e}vy noise, we reinterpreted the results of Ref. \cite{cohenRapid}.
Taking into account the superstatistical model, we conclude that some of
those results are not valid in general. Due to the interplay between the
noise statistics, the velocity of the power input, and the particle
dissipative dynamics, in the long time regime the probabilities of positive
and negative fluctuations of equal magnitude become identical. This result
follows from the asymptotic vanishing of the characteristic constant that
defines the work probability symmetry, Eqs. (\ref{LevyNormaQFR}) and (\ref%
{ZitaLevyVanishing}). On the other hand, the time-scaled work only satisfies
the generalized FR when its behavior is compatible with a LDP, that is, the
size of its characteristic fluctuations must to diminish with time (Fig. 2
and 3).

While the uniqueness of the present proposal was not proved, based on the
requirements that it satisfies \cite{unico}, one can conclude that it may be
considered as a valid and solid tool for analyzing nonequilibrium
fluctuations in systems characterized by power-law distributions. The
stretched relation with nonextensive thermodynamics \cite{TsallisBook}, as
well as its applicability in specific experimental setups \cite{NanoSuperFT}
are open problems that with certainty deserve extra analysis.

\section*{Acknowledgments}

This work was supported by CONICET, Argentina, under Grant No. PIP
11420090100211.

\appendix

\section{Large deviation functions}

In this Section it is shown the consistence of the proposed FR with a large
deviation-like principle. It is not obvious that an arbitrary generalized FR
may satisfies this condition. Specifically, we show that it is possible to
define two LDFs from the (long time) asymptotic behavior of the probability
density and its associated q-characteristic function. Both of them become
related by a Legendre-Fenchel transformation \cite{touchette}. These results
provide\ a solid mathematical support to the proposed FR.

\subsection{Unnormalized scheme}

We base our analysis on time-scaled variables, Eq. (\ref{rate}). A large
deviation-like principle relies on providing a general structure for the
probability $p(\mu )$\ in the long time regime. Instead of a standard
exponential structure \cite{touchette}, here we assume%
\begin{equation}
p(\mu )\asymp \frac{t^{\delta /2}}{C_{q}}\exp _{q}[-t^{\delta
}C_{q}^{1-q}\varphi (\mu )].  \label{AsymptoticP}
\end{equation}%
As before, the symbol $\asymp $ denotes an equality valid in a long time
regime. The factor $t^{\delta }$ $(\delta >0)$ measures the time-speed of $%
\varphi (\mu )\geq 0.$ As in Refs. \cite{klages,harris} (see also Appendix D
of Ref. \cite{touchette}), we consider the case in which $\delta \neq 1.$ On
the other hand, the factor $(t^{\delta /2}/C_{q})$ in front of the
q-exponential is necessary for providing the rights units and normalization
of $p(\mu ).$

After a simple manipulation without involving any extra approximation, Eq. (%
\ref{AsymptoticP}) can be rewritten\ as%
\begin{equation}
p(\mu )\asymp \frac{t^{\delta /2}}{C_{q}}\otimes _{q}\exp _{q}[-t^{\eta
}\varphi (\mu )],  \label{AsymptoticPExpor}
\end{equation}%
where the exponent $\eta $ reads%
\begin{equation}
\eta =\delta \Big{(}\frac{3-q}{2}\Big{)}.  \label{Neta}
\end{equation}%
Written is this way, given that $\lim_{t\rightarrow \infty }\ln
_{q}[t^{\delta /2}/C_{q}]/t^{\eta }=0,$ the function $\varphi (\mu )$ can be
obtained as%
\begin{equation}
\varphi (\mu )=\lim_{t\rightarrow \infty }\frac{-1}{t^{\eta }}\ln _{q}[p(\mu
)].  \label{phi}
\end{equation}%
Hence, it can be read as the probability's LDF \cite{touchette}.

Another LDF can be defined from the asymptotic time behavior of $%
Z_{q}(\lambda ).$ Its structure can be obtained from the definition (\ref%
{qCharacteristic}), after taking into account the probability asymptotic
behavior (\ref{AsymptoticP}). In general, the resulting integral cannot be
obtained exactly. Nevertheless, it can be worked out through a steepest
descent approximation. In Appendix B\ we derive the asymptotic expression%
\begin{equation}
Z_{q}(\lambda )\asymp \exp _{\tilde{q}}\Big{[}-\Big{(}\frac{3-q}{2}\Big{)}%
t^{\delta }C_{q}^{1-q}\Theta (\lambda )\Big{]},  \label{AsymptoticZetal}
\end{equation}%
where the index of the q-exponential reads%
\begin{equation}
\tilde{q}=\frac{1+q}{3-q}.  \label{qtilde}
\end{equation}%
Hence, the LDF associated to $Z_{q}(\lambda )$ can be defined as%
\begin{equation}
\Theta (\lambda )=\lim_{t\rightarrow \infty }\frac{-1}{t^{\delta }}\frac{2}{%
(3-q)}C_{q}^{q-1}\ln _{\tilde{q}}[Z_{q}(\lambda )].  \label{theta}
\end{equation}

The steepest descent approximation establishes a link between both LDF
(Appendix B). They are related by the Legendre-Fenchel transformation%
\begin{equation}
\Theta (\lambda )=\min_{\mu }[\varphi (\mu )+\lambda \mu ],
\label{ThetaLegendre}
\end{equation}%
jointly with the inverse equation%
\begin{equation}
\varphi (\mu )=\max_{\lambda }[\Theta (\lambda )-\lambda \mu ].
\label{phiLegendre}
\end{equation}%
These relations also arise from a standard LDP, where the asymptotic
behavior of the probability and its characteristic function scales with
standard exponential functions. Remarkably, this Legendre structure remains
valid even when the distributions develop power-law tails.

\subsubsection{Symmetries of the LDFs}

After establishing a large deviation-like principle [Eqs. (\ref%
{AsymptoticPExpor}) and (\ref{AsymptoticZetal})], we ask about the
symmetries that the LDFs must to satisfy when the generalized FR is valid in
the long time regime. A probability $p(\mu ),$ with the asymptotic structure
(\ref{AsymptoticPExpor}), satisfies the FR (\ref{FRProbaLong}) if the LDF $%
\varphi (\mu )$ fulfill the condition%
\begin{equation}
-\varphi (\mu )+\varphi (-\mu )=\alpha \mu .  \label{phiSymmetry}
\end{equation}%
Furthermore, the characteristic function (\ref{AsymptoticZetal}) satisfies
the symmetry (\ref{FRCaracLong}) if the LDF $\Theta (\lambda )$ satisfies%
\begin{equation}
\Theta (\lambda )=\Theta (-\lambda +\alpha ).  \label{thetaSymmetry}
\end{equation}%
Both conditions are consistent between them. In fact, one can be derived
from the other by using the Legendre structure defined by Eqs. (\ref%
{ThetaLegendre}) and (\ref{phiLegendre}). The demonstration is exactly the
same than in the standard case \cite{spohn,mukamel}.

\subsubsection{q-Gaussian distribution}

It is very instructive to exemplify the previous results with an arbitrary
q-Gaussian distributed variable. Let consider a stochastic variable $%
x_{st}(t)$ whose long time statistics is given by Eq. (\ref{qGaussiana})
under the replacements $x_{0}\rightarrow x_{0}(t)$ and $\beta \rightarrow
\beta (t).$ Furthermore, we assume that asymptotically these objects behave
as%
\begin{equation}
x_{0}(t)\asymp \mu _{0}t^{\kappa },\ \ \ \ \ \ \ \ \ \beta (t)\asymp \frac{%
\Delta _{0}}{t^{\kappa ^{\prime }}}.
\end{equation}%
Both, $\kappa $ and $\kappa ^{\prime }$ are positive exponents, while $\mu
_{0}$ and $\Delta _{0}$\ are characteristic constants. Notice that both the
average (strictly the most probable value) and the characteristic width of
the distribution $p(x)$ grows with time.

By defining $\mu _{st}(t)=x_{st}(t)/t^{\kappa }$ [Eq. (\ref{rate})], using
the change of measures $p(\mu )d\mu =p(x)dx,$ from Eq. (\ref{qGaussiana}) it
follows the distribution%
\begin{equation}
p(\mu )\asymp \frac{\sqrt{\Delta _{0}t^{2\kappa -\kappa ^{\prime }}}}{%
\mathcal{N}_{q}}\exp _{q}[-\Delta _{0}t^{2\kappa -\kappa ^{\prime }}(\mu
-\mu _{0})^{2}].  \label{PuQGauss}
\end{equation}%
Hence, by comparing with Eq. (\ref{AsymptoticP}) it follow the
identifications%
\begin{equation}
\delta =2\kappa -\kappa ^{\prime },\ \ \ \ \ \ \ \ \ C_{q}=\frac{\mathcal{N}%
_{q}}{\sqrt{\Delta _{0}}}.  \label{deltaCq}
\end{equation}%
The probability LDF reads%
\begin{equation}
\varphi (\mu )=\frac{(\Delta _{0})^{\frac{3-q}{2}}}{\mathcal{N}_{q}^{1-q}}%
(\mu -\mu _{0})^{2}.  \label{LDFPhiQGauss}
\end{equation}%
In order to be consistent with a LDP, the exponent $\delta $ must be
positive, $2\kappa >\kappa ^{\prime }.$ Hence, the width of the distribution 
$p(\mu )$ diminishes with time. This property is expected for a time-scaled
variable, Eq. (\ref{rate}). A common situation corresponds to $\kappa
=\kappa ^{\prime }=1$ giving $\delta =1.$ Even in this case, the exponent $%
\eta ,$\ Eq. (\ref{Neta}), which defines the limit (\ref{phi}), is different
from one for $q\neq 1.$

The q-characteristic function (\ref{qCharacteristic}) associated to the
distribution (\ref{PuQGauss}) can be obtained exactly by using the previous
result (\ref{ZQGauss}). After a simple change of variables, we get%
\begin{equation}
Z_{q}(\lambda )\asymp \exp _{\tilde{q}}\Big{\{}\Big{(}\frac{3-q}{2}\Big{)}%
t^{\delta }C_{q}^{1-q}\Big{[}\frac{\lambda ^{2}\mathcal{N}_{q}^{1-q}}{%
4(\Delta _{0})^{\frac{3-q}{2}}}-\lambda \mu _{0}\Big{]}\Big{\}},
\label{ZetalGaussMU}
\end{equation}%
where $\tilde{q}$ is given by Eq. (\ref{qTilde_qGauss}), $\delta $ and $%
C_{q} $\ by Eq. (\ref{deltaCq}). We notice that this structure corresponds
to that obtained from a steeps descent integration, Eq. (\ref%
{AsymptoticZetal}). In fact, that approximation is exact for a q-Gaussian
distribution.

By comparing Eq. (\ref{AsymptoticZetal}) and (\ref{ZetalGaussMU}), we obtain
the LDF $\Theta (\lambda ).$ It reads%
\begin{equation}
\Theta (\lambda )=-\frac{\lambda ^{2}\mathcal{N}_{q}^{1-q}}{4(\Delta _{0})^{%
\frac{3-q}{2}}}+\lambda \mu _{0}.  \label{LDFThetaQGauss}
\end{equation}

It is straightforward to prove that $\varphi (\mu )$ [Eq. (\ref{LDFPhiQGauss}%
)] and $\Theta (\lambda )$ [Eq. (\ref{LDFThetaQGauss})] are related by the
Legendre-Fenchel transformations (\ref{ThetaLegendre}) and (\ref{phiLegendre}%
). Furthermore, both LDF satisfy respectively the symmetries (\ref%
{phiSymmetry}) and (\ref{thetaSymmetry}) with the same constant $\alpha ,$
which reads%
\begin{equation}
\alpha =4\mu _{0}\frac{(\Delta _{0})^{\frac{3-q}{2}}}{\mathcal{N}_{q}^{1-q}}.
\label{AlfaMUqGauss}
\end{equation}%
As expected, when $q\rightarrow 1$ the expressions (\ref{LDFPhiQGauss}), (%
\ref{LDFThetaQGauss}), and (\ref{AlfaMUqGauss}) reduce to those
corresponding to a normal Gaussian distribution.

\subsection{Normalized Scheme}

Eqs. (\ref{phiSymmetry}) and (\ref{thetaSymmetry}) are equivalent to the
unnormalized FR (\ref{FRProbaLong}). The normalized FR (\ref{BiassedFR}) can
also be expressed through a set of renormalized LDFs. We define 
\begin{equation}
\varphi ^{ph}(\mu )=\lim_{t\rightarrow \infty }\frac{-1}{t^{\delta }}\Big{(}%
\int_{-\infty }^{\infty }d\mu \lbrack p(\mu )]^{q}\Big{)}\ln _{q}[p(\mu )],
\label{phiPhys}
\end{equation}%
and similarly%
\begin{equation}
\Theta ^{ph}(\lambda _{ph})=\lim_{t\rightarrow \infty }\frac{-1}{t^{\delta }}%
\ln _{\tilde{q}}[Z_{q}^{ph}(\lambda _{ph})].  \label{thetaPhys}
\end{equation}%
In the asymptotic regime, the relation (\ref{lambdaMU}) is equivalent to $%
\lambda _{ph}=\lambda \lim_{t\rightarrow \infty }\frac{1}{t^{\delta -\eta }}%
\int_{-\infty }^{\infty }d\mu \lbrack p(\mu )]^{q}.$ In fact, it is possible
to demonstrate that%
\begin{equation}
D_{q}^{1-q}\equiv \lim_{t\rightarrow \infty }\frac{1}{t^{\delta -\eta }}%
\int_{-\infty }^{\infty }d\mu \lbrack p(\mu )]^{q}=\Big{(}\frac{3-q}{2}%
\Big{)}C_{q}^{1-q},  \label{Dquqol}
\end{equation}%
where $(\delta -\eta )=(q-1)\delta /2.$ Hence, $\eta $ is the same exponent
defined in Eq. (\ref{Neta}). The last equality in Eq. (\ref{Dquqol}) follows
by calculating the integral through a steepest descent approximation, where $%
p(u)$ is given by (\ref{AsymptoticP}). In the derivation we used the result (%
\ref{normita}) and the equality $\frac{3-q}{2}=\Gamma (\frac{1}{q-1})\Gamma (%
\frac{1+q}{2(q-1)})/[\Gamma (\frac{q}{q-1})\Gamma (\frac{3-q}{2(q-1)})].$ By
comparing the LDFs corresponding to the unnormalized [Eqs. (\ref{phi}) and (%
\ref{theta})] and normalized [(\ref{phiPhys}) and (\ref{thetaPhys})]
schemes, from Eq. (\ref{Dquqol}) it follows the relations%
\begin{equation}
\varphi ^{ph}(\mu )=\frac{\varphi (\mu )}{D_{q}^{q-1}},\ \ \ \ \ \ \ \ \ \
\Theta ^{ph}(\lambda _{ph})=\frac{\Theta (\lambda _{ph}D_{q}^{q-1})}{%
D_{q}^{q-1}}.  \label{BiassedUnbiassed}
\end{equation}%
Taking into account Legendre-Fenchel transformations (\ref{ThetaLegendre})
and (\ref{phiLegendre}), Eq. (\ref{BiassedUnbiassed}) implies that%
\begin{equation}
\Theta ^{ph}(\lambda _{ph})=\min_{\mu }[\varphi ^{ph}(\mu )+\lambda _{ph}\mu
],  \label{PhysTeTa}
\end{equation}%
jointly with the inverse relation%
\begin{equation}
\varphi ^{ph}(\mu )=\max_{\lambda _{ph}}[\Theta ^{ph}(\lambda _{pj})-\lambda
_{ph}\mu ].  \label{PhysFi}
\end{equation}%
Therefore, the normalized definitions (\ref{phiPhys}) and (\ref{thetaPhys})
also maintain the Legendre structure associated to a large deviation theory.

\subsubsection{Symmetries of the LDFs}

The (unnormalized) symmetries (\ref{phiSymmetry}) and (\ref{thetaSymmetry}),
added to the\ relations (\ref{BiassedUnbiassed}), lead to the equivalent
relations%
\begin{equation}
-\varphi ^{ph}(\mu )+\varphi ^{ph}(-\mu )=\alpha ^{ph}\mu ,
\label{FR_PhysFI}
\end{equation}%
and consistently%
\begin{equation}
\Theta ^{ph}(\lambda _{ph})=\Theta ^{ph}(-\lambda _{ph}+\alpha ^{ph}),
\label{Fr_PhysTeta}
\end{equation}%
where the constant $\alpha _{ph}$\ reads%
\begin{equation}
\alpha _{ph}=\frac{\alpha }{D_{q}^{q-1}}.  \label{RelacionALFAS}
\end{equation}

\subsubsection{q-Gaussian distribution}

For the previous example, defined by the q-Gaussian distribution (\ref%
{PuQGauss}), it follows the LDF%
\begin{equation}
\varphi ^{ph}(\mu )=\Big{(}\frac{3-q}{2}\Big{)}\Delta _{0}(\mu -\mu
_{0})^{2}.
\end{equation}%
The q-characteristic function (\ref{ZetalGaussMU}) leads to%
\begin{equation}
Z_{q}^{ph}(\lambda _{ph})\asymp \exp _{\tilde{q}}\Big{\{}t^{\delta }\Big{[}%
\frac{\lambda _{ph}^{2}}{4(\frac{3-q}{2})\Delta _{0}}-\lambda _{ph}\mu _{0}%
\Big{]}\Big{\}}.
\end{equation}%
Therefore, we get%
\begin{equation}
\Theta ^{ph}(\lambda _{ph})=-\frac{\lambda _{ph}^{2}}{4(\frac{3-q}{2})\Delta
_{0}}+\lambda _{ph}\mu _{0}.
\end{equation}%
The constant $\alpha _{ph}$ reads%
\begin{equation}
\alpha _{ph}=4\mu _{0}\Delta _{0}\Big{(}\frac{3-q}{2}\Big{)}.
\end{equation}%
Notice that all dependences of the LDFs with the parameter $q$ can be
absorbed in the coefficient $\Delta _{0}.$

\section{Steepest descent approximation}

Here we develop a set of approximations that allow to calculate the long
time behavior of $Z_{q}(\lambda )$ from the asymptotic behavior of $p(\mu ).$
The procedure is similar to that given in Refs. \cite{mukamel} (see Appendix
C) and \cite{touchette}.

By introducing (\ref{AsymptoticP}) in Eq. (\ref{qCharacteristic}), after
some calculations steps, we get%
\begin{equation}
Z_{q}(\lambda )\asymp \frac{t^{\delta /2}}{C_{q}}\int_{-\infty }^{\infty
}d\mu \exp _{q}\{-t^{\delta }C_{q}^{1-q}[\varphi (\mu )+\lambda \mu ]\}.
\end{equation}%
This integral cannot be performed in an exact way. In order to proceed, we
introduce an integral representation of the q-exponential function $(q>1)$ 
\cite{TsallisBook,stepestQ}%
\begin{equation}
e_{q}^{-z}=\frac{1}{\Gamma (\frac{1}{q-1})}\int_{0}^{\infty }d\tau \tau ^{%
\frac{2-q}{q-1}}e^{-\tau \lbrack 1+(q-1)z]},  \label{IntegralRepresentation}
\end{equation}%
$(z>0).$ Hence, after inverting the order of the integrals, we write%
\begin{eqnarray}
Z_{q}(\lambda ) &\asymp &\frac{t^{\delta /2}}{\Gamma (\frac{1}{q-1})C_{q}}%
\int_{0}^{\infty }d\tau \tau ^{\frac{2-q}{q-1}}e^{-\tau }\int_{-\infty
}^{\infty }d\mu \\
&&\times \exp \{-t^{\delta }\tau \lbrack \varphi (\mu )+\lambda \mu
]C_{q}^{1-q}(q-1)\}.  \notag
\end{eqnarray}%
At long times $t,$ the integral in the $\mu $ variable can be worked out
with a steepest descent integration method. The main contribution to the
integral comes from the value of $\mu ,$ $\mu ^{\ast },$ that minimizes the
exponential. Defining $\mu ^{\ast }=\mu ^{\ast }(\lambda )$\ by the
condition $\dot{\varphi}(\mu ^{\ast })=-\lambda ,$ $[\dot{\varphi}%
(z)=(d/dz)\varphi (z)],$ we can approximate%
\begin{equation}
\varphi (\mu )+\lambda \mu \approx \varphi (\mu ^{\ast })+\lambda \mu ^{\ast
}+\frac{1}{2}\ddot{\varphi}(\mu ^{\ast })(\mu -\mu ^{\ast })^{2}.
\end{equation}%
By assuming that $\varphi (\mu )$ is a convex function to have a minimum, $%
\ddot{\varphi}(\mu ^{\ast })>0,$ and using the integral $\int_{-\infty
}^{\infty }d\mu e^{-a\mu ^{2}}=(\pi /a)^{1/2},$ it follows%
\begin{eqnarray}
Z_{q}(\lambda ) &\asymp &\frac{1}{\Gamma (\frac{1}{q-1})C_{q}^{\frac{3-q}{2}}%
}\frac{\sqrt{2\pi }}{\sqrt{(q-1)\ddot{\varphi}(\mu ^{\ast })}} \\
&&\times \int_{0}^{\infty }d\tau \tau ^{(\frac{2-q}{q-1}-\frac{1}{2}%
)}e^{\{-\tau \lbrack 1+t^{\delta }\Theta (\lambda )C_{q}^{1-q}(q-1)]\}}. 
\notag
\end{eqnarray}%
Here, $\Theta (\lambda )$ denote the function%
\begin{equation}
\Theta (\lambda )=\varphi (\mu )+\lambda \mu ,  \label{tetalLegendre}
\end{equation}%
where the value of $\mu $ follows from the condition%
\begin{equation}
\dot{\varphi}(\mu )=-\lambda .  \label{lambda}
\end{equation}%
By using again the integral representation (\ref{IntegralRepresentation}),
we arrive to the expression%
\begin{equation}
Z_{q}(\lambda )\asymp A_{q}\exp _{\tilde{q}}\Big{[}-\Big{(}\frac{3-q}{2}%
\Big{)}t^{\delta }C_{q}^{1-q}\Theta (\lambda )\Big{]},  \label{Zetolina}
\end{equation}%
where $\tilde{q}$ is given by Eq. (\ref{qtilde}), $\tilde{q}=(1+q)/(3-q).$
The constant $A_{q}$ is%
\begin{equation}
A_{q}\!=\!\frac{\Gamma (\frac{1}{\tilde{q}-1})\sqrt{2\pi }}{\Gamma (\frac{1}{%
q-1})C_{q}^{\frac{3-q}{2}}\sqrt{(q-1)\ddot{\varphi}(\mu ^{\ast })}}\!=\!%
\frac{\mathcal{N}_{q}}{C_{q}^{\frac{3-q}{2}}}\sqrt{\frac{2}{\ddot{\varphi}%
(\mu ^{\ast })}}\!\simeq \!1.  \label{normita}
\end{equation}%
Therefore, Eq. (\ref{Zetolina}) leads to the expression (\ref%
{AsymptoticZetal}). In the previous expression, the second equality follows
from Eq. (\ref{Nq}), while the last estimation follows by applying a
steepest descent approximation to the condition $\int_{-\infty }^{+\infty
}d\mu p(u)\asymp 1,$ where $p(u)$ is given by (\ref{AsymptoticP}), jointly
with the condition that the LDF $\varphi (\mu )$\ vanishes at its minimum $%
\mu _{\min },$ $\varphi (\mu _{\min })=0,$ which consistently implies $%
\Theta (0)=0,$ $Z_{q}(0)\asymp 1.$

Eqs. (\ref{tetalLegendre}) and (\ref{lambda}) show that $\Theta (\lambda )$
is the Legendre transform of $\varphi (\mu ),$ which in turn can be written
as in Eq. (\ref{ThetaLegendre}). On the other hand, the derivative of $%
\Theta (\lambda )$ with respect to $\lambda $ is%
\begin{equation}
\dot{\Theta}(\lambda )=\dot{\varphi}(\mu )\frac{d\mu }{d\lambda }+\mu
+\lambda \frac{d\mu }{d\lambda },
\end{equation}%
which from Eq. (\ref{lambda}) leads to%
\begin{equation}
\dot{\Theta}(\lambda )=\mu .  \label{utilde}
\end{equation}%
This shows that $\varphi (\mu )$ is given by the inverse Legendre transform
of $\Theta (\lambda ),$%
\begin{equation}
\varphi (\mu )=\Theta (\lambda )-\lambda \mu .
\end{equation}%
Here the value of $\lambda $ follows from the condition (\ref{utilde}),
leading to Eq. (\ref{phiLegendre}). In fact, by taking the derivative of (%
\ref{utilde}) with respect to $\lambda $ and using the derivative of (\ref%
{lambda}) with respect to $\mu $ we can confirm that $\varphi (\mu )$ is
convex because $\Theta (\lambda )$ is concave, $\ddot{\varphi}(\mu )=-1/%
\ddot{\Theta}(\lambda ).$


\begin{thebibliography}{99}
\bibitem{gallavoti} G. Gallavotti and E.G.D. Cohen, Phys. Rev. Lett. \textbf{%
74}, 2694 (1995); D.J. Evans, E.G.D. Cohen, and G.P. Morris, Phys. Rev.
Lett. \textbf{71}, 2401 (1993).

\bibitem{Jarzynski} C. Jarzynski, Phys. Rev. Lett. \textbf{78}, 2690 (1997).

\bibitem{kurchan} J. Kurchan, J. Phys. A \textbf{31}, 3719 (1998).

\bibitem{spohn} J.L. Lebowitz and H. Spohn, J. Stat. Phys. \textbf{95}, 333
(1999).

\bibitem{crocks} G.E. Crocks, Phys. Rev. E \textbf{60}, 2721 (1999).

\bibitem{maes} C. Maes, J. Stat. Phys. \textbf{95}, 367 (1999).

\bibitem{seifert} U. Seifert, Phys. Rev. Lett. \textbf{95}, 040602 (2005).

%Reviews on fluctuation theorem

\bibitem{searles} E.M. Sevick, R. Prabhakar, S.R. Williams, and D.J.
Searles, Annu. Rev. Phys. Chem \textbf{59}, 603 (2008).

\bibitem{vulpiani} U.M.B. Marconi, A. Puglisi, L. Rondoni, and A. Vulpiani,
Phys. Rep. \textbf{461}, 111 (2008).

\bibitem{mukamel} M. Esposito, U. Harbola, and S. Mukamel, Rev. Mod. Phys. 
\textbf{81}, 1665 (2009).

\bibitem{JarzynskiReport} C. Jarzynski, Annu. Rev. Condens. Matter Phys. 
\textbf{2}, 329 (2011).

%FT Experiments

\bibitem{PRLexper} %P. Gaspard, J. Chem. Phys. \textbf{120}, 8898 (2004);
W.I. Goldburg, Y.Y. Goldschmidt, and H. Kellay, Phys. Rev. Lett. \textbf{87}%
, 245502 (2001); D.M. Carberry, J.C. Reid, G.M. Wang, E.M. Sevick, D.J.
Searles, and D.J. Evans, Phys. Rev. Lett. \textbf{92}, 140601 (2004); A.
Puglisi, P. Visco, A. Barrat, E. Trizac, and F. van Wijland, Phys. Rev.
Lett. \textbf{95}, 110202 (2005); S. Schuler, T. Speck, C. Tietz, J.
Wrachtrup, and U. Siefert, Phys. Rev. Lett. \textbf{94}, 180602 (2005); M.
Belushkin, R. Livi, and G. Foffi, Phys. Rev. Lett. \textbf{106}, 210601
(2011).%
% J. Mehl, T. Speck, and U. Seifert, Phys. Rev. E \textbf{78}, 011123 (2008).

\bibitem{soodPituto} A.A. Budini, Phys. Rev. E \textbf{84}, 061118 (2011);
N. Kumar, S. Ramaswamy, and A.K. Sood, Phys. Rev. Lett. \textbf{106}, 118001
(2011).

%Standard Work fluctuation theorem with white noise

\bibitem{mazonka} O. Mazonka and C. Jarzynski, arXiv:cond-mat/9912121 (1999).

\bibitem{zon} R. van Zon and E.G.D. Cohen, Phys. Rev. E \textbf{67}, 046102
(2003).

%ExperimentoBrownianoFT

\bibitem{wangBrownianExp} G.M. Wang, E.M. Sevick, E. Mittag, D.J. Searles,
and D.J. Evans, Phys. Rev. Lett. \textbf{89}, 050601 (2002).

%Extension fluctuation theorem (Non-linear) for heat

\bibitem{FTExtended} R. van Zon and E.G.D. Cohen, Phys. Rev. Lett. \textbf{91%
}, 110601 (2003); Phys. Rev. E \textbf{69}, 056121 (2004).

%FT with Poisson noise

\bibitem{cohenPoisson} A. Baule and E.G.D. Cohen, Phys. Rev. E \textbf{79},
030103(R) (2009); Phys. Rev. E, \textbf{80}, 011110 (2009).

%Extension fluctuation theorem (Non-linear) for noise driven systems

\bibitem{FTNonLinear} J.R. Gomez-Solano, L. Bellon, A. Petrosyan, and S.
Ciliberto, EPL \textbf{89}, 60003 (2010); M. Bonaldi, \textit{et. al.},
Phys. Rev. Lett. \textbf{103}, 010601 (2009); C. Falcon and E. Falcon, Phys.
Rev. E \textbf{79}, 041110 (2009); E. Falcon, S. Aumaitre, C. Falcon, C.
Laroche, and S. Fauve, Phys. Rev. Lett. \textbf{100}, 064503 (2008); Jean
Farago, J. Stat. Phys. \textbf{107}, 781 (2002).

%Large deviation theory

\bibitem{touchette} H. Touchette, Phys. Rep. \textbf{478}, 1 (2009).

\bibitem{sornette} D. Sornette, \textit{Critical Phenomena in Natural
Sciences}, (Springer, 2006).%
%\bibitem{PointProcesses} S.B. Lowen and M.C. Teich, \textit{Fractal-Based
%Point Processes}, (Wiley\&Sons, 2005).

\bibitem{garrahan} L.O. Hedges, R.L. Jack, J.P. Garrahan, and D. Chandler,
Science \textbf{323}, 1309 (2009); J.P. Garrahan and I. Lesanovsky, Phys.
Rev. Lett. \textbf{104}, 160601 (2010); J.P. Garrahan, A.D. Armour, and I.
Lesanovsky, Phys. Rev. E \textbf{84}, 021115 (2011); A.A. Budini, Phys. Rev.
E \textbf{82}, 061106 (2010); Phys. Rev. E \textbf{84}, 011141 (2011).

%Levy and work fluctuation theorem

\bibitem{cohenRapid} H. Touchette and E.G.D. Cohen, Phys. Rev. E \textbf{76}%
, 020101(R) (2007); Phys. Rev. E \textbf{80}, 011114 (2009).

\bibitem{klages} A.V. Chechkin and R. Klages, J. Stat. Mech.: Theory Exp.
(2009), L03002.

%CurrentFluctuationsNonLinear Scaling in FR

\bibitem{harris} R.J. Harris and H. Touchette, J. Phys. A \textbf{42},
342001 (2009).

%FT in superstatistics

\bibitem{BeckCohenFT} C. Beck and E.G.D. Cohen, Phys. A \textbf{344}, 393
(2004).

%SuperStatistics

\bibitem{beckQ} C. Beck, Phys. Rev. Lett. \textbf{87}, 180601 (2001).

\bibitem{superstatistics} C. Beck and E.G.D. Cohen, Phys. A \textbf{322},
267 (2003); H. Touchette and C. Beck, Phys. Rev. E \textbf{71}, 016131
(2005); S. Abe, C. Beck, and E.G.D. Cohen, Phys. Rev. E \textbf{76}, 031102
(2007).

%NonExtensive Thermodynamics

\bibitem{TsallisBook} C. Tsallis, \textit{Introduction to Nonextensive
Statistical Mechanics}, (Springer, 2009).

\bibitem{Qlectures} M. Sugiyama, ed., \textit{Nonadditive Entropy and
Nonextensive Statistical Mechanics}, Continuum Mechanics and Thermodynamics 
\textbf{16} (Springer-Verlag, Heidelberg, 2004); P. Grigolini, C. Tsallis,
and B.J. West, eds., \textit{Classical and Quantum Complexity and
Nonextensive Thermodynamics}, Chaos, Solitons and Fractals \textbf{13},
Issue 3 (2002); S. Abe and Y. Okamoto, eds., \textit{Nonextensive
Statistical Mechanics and its Applications}, Series \textit{Lecture Notes in
Physics} \textbf{560} (Springer, Berlin, 2001).%
%Europhysicsnews \textbf{36/6} (2005),  
%\textit{Nonextensive statistical mechanics: new trends, new perspectives}.

\bibitem{QCentralLimit} S. Umarov, C. Tsallis, and S. Steinberg, Milan J.
Math. \textbf{76}, 307 (2008); A. Rodriguez, V. Schw\"{a}mmle, and C.
Tsallis, J. Stat. Mech.: Theory Exp. (2008), P09006; R. Hanel, S. Thurner,
and C. Tsallis, Eur. Phys. J. B \textbf{72}, 263 (2009).

\bibitem{QFokker} C. Tsallis and D.J. Bukman, Phys. Rev. E \textbf{54},
R2197 (1996); M. Bologna, C. Tsallis, and P. Grigolini, Phys. Rev. E \textbf{%
62}, 2213 (2000).

\bibitem{qConstraints} C. Tsallis, R.S. Mendes, and A.R. Plastino, Phys. A 
\textbf{261}, 534 (1998).

\bibitem{LevyAsQ} D. Prato and C. Tsallis, Phys. Rev. E \textbf{60}, 2398
(1999); C. Tsallis, S.V.F. Levy, A.M.C. Souza, and R. Maynard, Phys. Rev.
Lett. \textbf{75}, 3589 (1995).

\bibitem{stepestQ} S. Abe and A.K. Rajagopal, J. Phys. A \textbf{33}, 8733
(2000).

\bibitem{NanoSuperFT} V. Garcia-Morales and K. Krischer, Proc. Natl. Acad.
Sci. U.S.A. \textbf{108}, 19535 (2011).%
%\bibitem{vanKampen} N.G. van Kampen, \textit{Stochastic Processes in Physics
%and Chemistry}, (Sec. Ed., North-Holland, Amsterdam, 1992).

\bibitem{levyNum} R.H. Rimmer and J.P. Nolan, Math. J. \textbf{9}, 776
(2005).

\bibitem{unico} Different generalizations of the logarithmic function \cite%
{kania} may be the basis for proposing alternative FRs. Nevertheless, it is
not clear at this point if any alternative definition can satisfy the
properties of the present approach: (i) The FR must be expressed by a linear
dependence when power-law arises. (ii) It must be possible to define the
symmetry in terms of a characteristic-like function. (iii) The long time
behavior of the probability and its characteristic function must be defined
in terms of a set of LDF's related by a Legendre transform.

\bibitem{kania} P. Tempesta, Phys. Rev. E \textbf{84}, 021121 (2011); S.
Umarov, C. Tsallis, M. Gell-Mann, and S. Steinberg, J. Math. Phys. \textbf{51%
}, 033502 (2010).
\end{thebibliography}
\end{document}